\documentclass{aa}

\usepackage{graphicx}
\usepackage{subcaption}         
\usepackage{lscape}             
\usepackage{placeins}           
                                
\usepackage{natbib}
\usepackage[hidelinks]{hyperref}
\hypersetup{
     colorlinks   = true,
     linkcolor    = blue,
     citecolor    = blue
}

\usepackage[T1]{fontenc}

\usepackage{graphicx}	
\usepackage{amsmath}	
\usepackage{amssymb}	
\newcommand{\be}{\begin{equation}}
\newcommand{\ee}{\end{equation}}
\newcommand{\beal}{\begin{aligned}}
\newcommand{\eeal}{\end{aligned}}
\usepackage{xcolor}

\begin{document}
\title
{
A simple toy model for the electromagnetic variability of lump-dominated circumbinary disks around binary black holes
}
\titlerunning{A toy model for the EM variability of CBDs around BBHs}
\authorrunning{Mignon-Risse et al.}
\author{
Rapha\"{e}l Mignon-Risse,$^{1,2,3,4}$\thanks{E-mail: raphael.mignon-risse@lam.fr}
Peggy Varniere,$^{2,5}$
Fabien Casse$^{2}$}

\institute{Department of Physics, Norwegian University of Science and Technology, NO-7491 Trondheim, Norway
\and Universit\'e Paris Cit\'e, CNRS, Astroparticule et Cosmologie, F-75013 Paris, France
\and Department of Theoretical Physics, Atomic and Optics, Campus Miguel Delibes, University of Valladolid, Paseo Bel\'en, 7, 47011, Valladolid, Spain
\and Aix-Marseille Universit\'e, CNRS, CNES, LAM, Marseille, France
\and Universit\'e Paris-Saclay, Universit\'e Paris Cit\'e, CEA, CNRS, AIM, 91191, Gif-sur-Yvette, France
}

\date{Accepted XXX. Received YYY; in original form ZZZ}

\abstract
   {The electromagnetic detection of circumbinary disks around pre-merger binary black holes (BBHs) relies on theoretical predictions.
These are generally obtained through expensive numerical simulations, but simple or fast toy models are lacking to unleash the potential of these theoretical advances for observational purposes.}
{We aim to present a simple toy model to compute the electromagnetic variability of circumbinary disks around circular-orbit BBHs {at relativistic} separations, focusing on the impact of disk non-axisymmetries.}
{{We assume that the disk is threaded by spiral arms and hosts a hotspot linked to an overdense structure (the {\lq}lump{\rq}) preferably reported in binaries close to equal mass.}
 We build a simple temperature distribution, and estimate its thermal emission, perceived by a distant observer, via a ray-tracing code in a BBH approximate metric.
We propose a toy model reproducing the main lightcurve features and show it is consistent with 2D general-hydrodynamical simulations {under the assumption of compressional heating and expansional cooling except for purely dynamical effects such as the binary-lump beat}.
}
{The lightcurve exhibits {a main modulation }at the lump's period (i.e. a few times the orbital period), due to relativistic Doppler effect{, and a shorter one at the orbital-like period, due to spiral arms or the beat.}
These are more prominent in the optical/UV band for a total binary mass $M\, {=} \, 10^{4-10}\mathrm{M_\odot}$, where the disk energy spectrum peaks.
{For $M=10^{9}\mathrm{M_\odot}$, a $4\%$-amplitude lump modulation is detectable with the Vera Rubin Observatory after six months of observation, up to $z\, {=}\, 0.5$.}
}
{We proposed a new, simple toy model that can be used, for instance, to test the compatibility of the periodicity of BBH candidate sources with a circumbinary disk origin.
}

\maketitle



\section{Introduction}

Supermassive binary black holes (SMBBHs) are believed to emit copious amount of electromagnetic (EM) radiation 
prior to merger\footnote{{As the inspiral motion accelerates, the BBH could {\lq}decouple{\rq} from the circumbinary disk (\citealt{milosavljevic_afterglow_2005}, \citealt{dittmann_decoupling_2023}), 
affecting its EM appearance (\citealt{krauth_disappearing_2023}, \citealt{franchini_emission_2024}).}} and their future EM observation is a milestone of the multi-messenger era.
Indeed, individual SMBHs reside at the center of their host galaxies (e.g. \citealt{ferrarese_supermassive_2005}, \citealt{gultekin_fundamental_2009}), {so} merging galaxies should provide a gas-rich environment for the SMBBH.
{Emitting lower-frequency gravitational waves (GWs) than the now widely detected stellar-mass BBHs (\citealt{abbott_observation_2016}, \citealt{abbott_gwtc-2_2021}, 
\citealt{the_ligo_scientific_collaboration_population_2022}), th}e most massive SMBBHs ({total mass} $M\,{>}\,10^8\,\mathrm{M_\odot}$) are now detectable with Pulsar Timing Arrays (PTA, e.g. 
\citealt{lommen_pulsar_2015}, \citealt{babak_european_2016}, \citealt{kelley_multi-messenger_2019}) in the $1$~nHz$-1 \, \mathrm{\mu Hz}$ band while BBHs with $M\, {=} \, 10^{4-7}\mathrm{M_\odot}$ 
will be detectable with the Laser Interferometer Space Antena (LISA, \citealt{amaro-seoane_laser_2017}) in the $0.1$~mHz$\, {-}\,1$~Hz band.
Interestingly, SMBBH GW detection with LISA could be possible days to months before the merger (see \citealt{mangiagli_observing_2020}) thus leaving room for pre-merger EM follow-up.
LISA's error box can be as large as $10-1000\, \mathrm{deg}^2$, {with} numerous sources to search through to identify the origin of the GW signal.
For that reason{, numerous} efforts have been undertaken to identify the tell-tale EM signal of BBHs and search for candidates
 (\citealt{liu_supermassive_2019}, \citealt{de_rosa_quest_2019}, \citealt{dorazio_observational_2023}).
 The present paper is meant to support such efforts.

Studies dedicated to predict{ing} such pre-merger EM signature{s} rely on analytical approaches and/or (magneto-~)hydrodynamical simulations, deriving from a common agreement on the flow geometry when the mass ratio $q$ of the binary is not too extreme (typically ${\sim} \, 0.1\, {-}\, 1$).
For aligned, circular binaries, it consists in the following:
a cavity {of} about twice the orbital separation \citep{artymowicz_dynamics_1994} because no stable circular orbit exists {there}; two spiral waves threading the circumbinary disk (CBD; e.g. \citealt{macfadyen_eccentric_2008}) and spiral streams {penetrating the cavity} \citep{artymowicz_mass_1996} and eventually feed{ing} individual accretion structures referred to as \lq mini-disks{\rq} around the BHs (e.g. \citealt{farris_binary_2014}); finally, an overdensity dubbed \lq lump{\rq} \citep{shi_three-dimensional_2012} orbiting at the inner edge of the {CBD.
In this context, our aim is to compute the EM observables coming exclusively from the CBD around BBHs}, an expected configuration if gas is available (e.g. \citealt{artymowicz_mass_1996}) at a separation compatible with LISA detection (\citealt{piro_athena_2022} and references therein). 
{As we focus on the variability from the CBD we omit for now, on purpose, any other emission coming from possible mini-disks (e.g. \citealt{farris_characteristic_2015}, \citealt{dascoli_electromagnetic_2018}, \citealt{gutierrez_electromagnetic_2022}, \citealt{krauth_disappearing_2023}, \citealt{franchini_emission_2024}, \citealt{cocchiararo_electromagnetic_2024}, \citealt{porter_parameter_2024}), whose contribution would add up to the CBD's emission.} 
{In our model, we initially assume a non-axisymmetric temperature distribution in the CBD with perturbations due to the lump and spiral arms.}

The lump feature {has been reported in a large set of} CBD simulations {{around nearly-}circular (eccentricity $e\, {\lesssim} \, 0.1$, e.g. \citealt{dorazio_fast_2024}) binaries {with mass ratios of order unity}}.
{This \lq lump\rq{}  \citep{shi_three-dimensional_2012}  term} characterizes the presence of an asymmetry in the density (or $m=1$ azimuthal mode{, with $m$ the mode number)} located close to the CBD inner edge, orbiting at ${\sim}4-10\, \mathrm{P_{orb}}$ (e.g. \citealt{shi_three-dimensional_2012}, \citealt{mignon-risse_origin_2023}), with $\mathrm{P_{orb}}$ the binary period. 
Hence, this term does not refer to the underlying mechanism but to a feature from CBDs that was unexpected, particularly for initially symmetric ($q\, {=} \, 1$) systems. 
The interest for the lump comes from the intuitive prediction that it may, in time, modulate  the luminosity of CBDs 
as it orbits (see e.g. \citealt{gold_relativistic_2019}) and/or modulates the accretion rate (e.g. \citealt{munoz_circumbinary_2020}), but also that its associated gravitational torque may affect the orbital evolution of the central binary \citep{tiede_gas-driven_2020}.
The lump has been reported in 2D (e.g. \citealt{macfadyen_eccentric_2008}, \citealt{dorazio_accretion_2013}, \citealt{munoz_pulsed_2016}, \citealt{miranda_viscous_2017}, \citealt{tang_orbital_2017}, \citealt{mosta_gas_2019}, \citealt{munoz_hydrodynamics_2019}, \citealt{tiede_gas-driven_2020}, \citealt{tiede_how_2022}, \citealt{westernacher-schneider_multiband_2022}, \citealt{siwek_preferential_2022}, \citealt{krauth_disappearing_2023}, \citealt{mahesh_analytical_2023}, \citealt{wang_role_2023}) as well as three-dimensional (3D) simulations.
Those include grid-based (\citealt{moody_hydrodynamic_2019}, \citealt{shi_three-dimensional_2015}) as well as smoothed-particle hydrodynamic (\citealt{ruge_structures_2015}, \citealt{heath_orbital_2020}, \citealt{ragusa_evolution_2020}, \citealt{franchini_resolving_2022}, \citealt{franchini_emission_2024}) studies.
Regarding the angular momentum transport mechanism, the lump was found in non-viscous (e.g. \citealt{mignon-risse_origin_2023}, \citealt{cimerman_gravitational_2023}), viscous (e.g. \citealt{farris_binary_2014}, \citealt{farris_characteristic_2015}) hydrodynamical as well as MHD (e.g. \citealt{shi_three-dimensional_2012}) and radiation-MHD \citep{tiwari_radiation_2025} simulations.
It was also present regardless of the gravity description: Newtonian (e.g. \citealt{dittmann_survey_2022}), Post-Newtonian (\citealt{liu_evolution_2021}) or GR (e.g. \citealt{noble_circumbinary_2012}, \citealt{gold_accretion_2014}, \citealt{zilhao_resolving_2015}, \citealt{noble_mass-ratio_2021}, \citealt{lopez_armengol_circumbinary_2021}, \citealt{mignon-risse_origin_2023}).
{For circular binaries,} it is found to form {when $q$ is not too far from unity ($q\gtrsim 0.25$, \citealt{dorazio_accretion_2013}; $q\gtrsim 0.1$, \citealt{mignon-risse_origin_2023}; $q\gtrsim 0.5$, \citealt{noble_mass-ratio_2021}).}
Since the orbital separation can be and was used to renormalize distances in {most} Newtonian studies (e.g. \citealt{miranda_viscous_2017}), the lump could be a common feature at all orbital separations allowing for the presence of a CBD.
Hence, those works suggest that the lump is a robust feature from CBDs orbiting circular binaries of comparable masses.

Moreover, {CBDs exhibiting a lump are likely to possess a non-axisymmetric temperature distribution with $m\, {=} \, 1$ symmetry
, as found by e.g. \citealt{tang_late_2018}, \citealt{gutierrez_electromagnetic_2022}, \citealt{cocchiararo_electromagnetic_2024}, \citealt{tiwari_radiation_2025}.
Possible causes are the enhanced shocks there (\citealt{shi_three-dimensional_2012},  \citealt{cocchiararo_electromagnetic_2024}), compressional heating, viscous heating (in the general sense, related to angular momentum transport, \citealt{cimerman_gravitational_2023}; see also the radial mass flux in \citealt{shi_three-dimensional_2012}), or the indirect impact of a strongly enhanced density on the disk thermodynamics.
\cite{wang_hydrodynamical_2023} found the temperature distribution to be axisymmetric only in their locally-isothermal CBDs, whereas it is non-axisymmetric for any other choice of the so-called $\beta-$cooling timescale.
Quantitatively, the temperature perturbation would naturally depend on the CBD thermodynamics, which has been modeled in several ways (see the references above) and is not well constrained at this stage.
Hence, we choose to compute the temperature perturbations via a useful parametric approach encapsulating these uncertainties.} 
In complement, {the lump} was proposed to be possibly linked to the Rossby Wave Instability \citep{mignon-risse_origin_2023}{, a well-studied instability (e.g. \citealt{tagger_accretionejection_2006,lovelace_rossby_2014}) known to develop at the edge of accretion disks \citep{lovelace_rossby_1999}}.
Thus, we can use these theoretical {(knowing the instability criterion)} or empirical arguments to extrapolate not only the spiral wave-related but also the lump-related emission modulation properties{, at least qualitatively}: this is the purpose of the present paper. 

{It is expected} that the Spectral Energy Distribution (SED) of the CBD peaks in the UV (\citealt{roedig_observational_2014}, \citealt{farris_characteristic_2015}, \citealt{dascoli_electromagnetic_2018}, \citealt{gutierrez_electromagnetic_2022}, \citealt{krauth_disappearing_2023}, \citealt{franchini_emission_2024}) to soft X-ray \citep{tang_late_2018} for $M\,{=}\,10^6\, \mathrm{M_\odot}$.
For non-face-on observers, any aforementioned non-axisymmetry orbiting in the disk would produce an EM modulation (e.g. \citealt{varniere_flux_2005}) at its orbital period, as a consequence of self-shadowing, further amplified by relativistic effects.

General-relativistic (GR) effects can be decisive to compute the EM variability.
Indeed, they modify in various aspects the observational appearance of disks around compact objects (e.g. relativistic Doppler, gravitational beaming, time delay), even for unresolved sources (e.g. \citealt{vincent_flux_2013}).
In the context of BBHs, {relativistic Doppler {(see e.g. \citealt{dorazio_relativistic_2015} on top of mini-disk models)} and gravitational beaming were accounted for in} \cite{tang_late_2018} on top of {pseudo-}Newtonian simulations, and by \cite{dascoli_electromagnetic_2018}, \cite{gutierrez_electromagnetic_2022}, \cite{porter_parameter_2024} with a BBH metric.
{Here, we consider these GR effects by using} an approximate BBH metric. 
{Since the CBD inner edge is at a few times the separation, we can expect relativistic effects -- which are more intense in the vicinity of BHs -- to be amplified at smaller separation; thus, we focus on relativistic separations (${\lesssim}\, 100 \, \mathrm{G}M/\mathrm{c}^2$) here.
At these separations, the binary should have already circularized efficiently through GW emission to reach an eccentricity $e\, {\lesssim} \, 0.1$ compatible with the lump's presence \citep{dorazio_fast_2024}, as discussed in our Appendix~\ref{app:ecc} (see also \citealt{zrake_equilibrium_2021}).
Accordingly, we will consider BBHs systems with circular orbits.}

The remainder of this paper is structured as follows.
In Sec.~\ref{sec:gyo}, we present the CBD model and ray-tracing approach.
In Sec.~\ref{sec:LC}, we compute and characterize the thermal lightcurve (LC) variability owing to the CBD's spiral arms and lump.  
Then, we propose a simple toy model for the LCs (Sec.~\ref{sec:toy}).
{W}e show it compares well with a LC produced from post-processed 2D general-hydrodynamical simulation{s} (Sec.~\ref{sec:lc}) {but in some cases the second modulation can be at the beat frequency between the binary and the lump}. 
We discuss how this toy model can be used {and apply it to searches with the Vera Rubin Observatory} in Sec.~\ref{sec:discu}{. We present} our main conclusions in Sec.~\ref{sec:ccl}.
\newline

In the following, we use the unit system $\mathrm{G}\, {=}\, \mathrm{c}\, {=}\, 1$, so that 
the gravitational radius, $r_\mathrm{g} \, {=}\,  \mathrm{G}M/\mathrm{c}^2$, which is the natural length unit, and the associated time unit $t_\mathrm{g} \, {=}\,  \mathrm{G}M/\mathrm{c}^3$, are simply equal to $r_\mathrm{g} \, {=}\,  t_\mathrm{g} \, {=}\,  \mathrm{M}$.
In that system, spatial lengths and times are given in units of $\mathrm{M}$ with the conversion factor $1\, \mathrm{M_\odot} \, {=}\,  1.477\,  \mathrm{km} \, {=}\,  4.926 \times 10^{-6} \, \mathrm{s}$.

\section{Simple representation of CBD characteristics}
\label{sec:gyo}

 In this paper we aim to explore {some of the thermal emission} characteristics of near-circular BBHs{'} CBD{s}. 
 Rather than doing the full GR-ray-tracing on {many} numerical simulations of CBDs to {cover the vast parameter space (in mass, mass ratio, separations, accretion rate...)}
 , we decided to start with a simplified, parametric, model exhibiting the key CBD characteristics found across {most of the simulations of near-circular BBHs in the literature}.
 {The model is} used to compute {the frequency- and time-dependent emission in GR.}
 The aim is to {identify potential key observables} and their relation to the binary parameters.

\subsection{Key CBD characteristics}
\label{sec:key}

As already presented in the introduction, the CBDs key structures consist in the following: 

\begin{description}
    \item[\tt -] a disk with an edge around twice the binary separation   \citep{roedig_observational_2014,farris_characteristic_2015,dascoli_electromagnetic_2018,
    gutierrez_electromagnetic_2022,krauth_disappearing_2023,franchini_emission_2024} {that we assume, for simplicity, to be symmetric.}   
  
    \item[\tt -]  {on top of the axisymmetric disk, we consider the so-called ${m\, {=} \, 1}$} \lq lump\rq\ consisting in an overdensity orbiting near the edge of the CBD {related to the local orbital period there}
    
    \item[\tt -] {finally we also add the} { ${m\, {=} \, 2}$} {(}two-arms{)} spiral{ embedded in the CBD} linked with the orbital period of the binary \citep{tang_late_2018,dascoli_electromagnetic_2018,duffell_circumbinary_2020,
    westernacher-schneider_multiband_2022,cimerman_gravitational_2023,franchini_emission_2024,cocchiararo_electromagnetic_2024}.

\end{description}

\subsection{{Temperature distribution model}}
\label{sec:repr_cbd}

 In order to mimic the key characteristics of the CBD we are using a similar method as in \citet{varniere_impact_2016} 
 where we have a{n} axisymmetric equilibrium disk onto which both the spiral and the lump are considered perturbations.  
From this complete temperature profile we can then compute the multicolor disk blackbody emission.

   {T}o represent the unperturbed, axisymmetric thermal emission {component} of the CBD we use the standard model first presented by  \cite{shakura_black_1973}
   {consisting in} a geometrically{-}thin, optically-thick disk {truncated} at an inner radius $r_{\mathrm{in}} \, {=} \, 2 \, \mathrm{b}$ {with $\mathrm{b}$ the separation}
    (as in \citealt{roedig_observational_2014}),
   with a monotonic temperature profile 
 \begin{eqnarray}
T_0(r) \simeq 2.4 \times 10^{6} \left( \frac{\dot{M}}{ \mathrm{\dot{M}}_\mathrm{Edd}} \left( \frac{4 \times 10^6 M_\odot}{M}\right) \left( \frac{r}{r_\mathrm{g}} \right)^{-3} \right)^\frac{1}{4}  \mathrm{K} 
\label{eq:T}
 \end{eqnarray}
with $\mathrm{\dot{M}}_\mathrm{Edd}$ the Eddington accretion rate, and a standard radiative efficiency $\eta\, {=} \,0.1$.

   To add a hotter spiral and lump on top of the {axisymmetric} disk, we use the {perturbation's function} presented in  \citet{varniere_impact_2016} which was successfully used to 
   produce synthetic observations of 
   {accretion disks around BHs of various masses} 
   \citep{varniere_possible_2016,varniere_reproducing_2017,varniere_quasiperiodic_2023}.
   {The temperature is given as a function of time and position as}
 \begin{eqnarray}
   \label{eq:Tna}
 T(t,r,\varphi)  \, {=} \, T_0(r) 
 \left[  1 + \gamma_\mathrm{s/l} \left(\frac{r_\mathrm{c}}{r}\right)^{1/4}\ e^{\left(-\frac{1}{2}\left(\frac{r-r_\mathrm{s/l}(t,r,\varphi)}{ \delta \,  r_\mathrm{c}}\right)^2\right)}\right]^2 
  \end{eqnarray}
  
  where  {the amplitude of the temperature perturbation due to} the non-axisymmetric pattern {is controlled by $\gamma_\mathrm{s}$ for the spiral and $\gamma_\mathrm{l}$ for the lump. 
  A choice of $\gamma_\mathrm{l} \, {=} \, 0.3$, the largest value we will consider here, leads to a temperature enhancement of $T/T_0 \, {\sim} \, 1.7$.
  In comparison, the simulations of \cite{tang_late_2018} show a contrast in effective temperature between the lump and its surroundings of a factor of $T/T_0 \, {\sim}\, 2$ or more (see their Fig.~1, bottom panel, with differences visible despite the logarithmic scale).
  The simulations of \cite{noble_mass-ratio_2021} do not show temperature maps but exhibit surface density constrasts of ${\approx} \, 2$ (see their Fig.~13, 14). 
  For simplicity, assuming only adiabatic and reversible heating so $P \, {\varpropto} \, \Sigma^{5/3}$, and $T \, {=} \, P/\Sigma$ (ideal gas), this translates into a temperature perturbation of $T/T_0 \, {\sim} \, 1.6$.
  Both are roughly consistent with the highest value of $\gamma_\mathrm{l}$ considered here. 
  With a similar reasoning based on the spiral density maps in the 2D hydrodynamical simulations of \cite{cimerman_gravitational_2023}, we will take $\gamma_\mathrm{s}$ such that $T/T_0 \lesssim 1.1$ (for $\gamma_\mathrm{l}\, {=} \, 0$).
  The pattern radial extent is controlled by $\delta$ which is fixed to $0.2$ for both the lump and the spiral.} 
      {The pattern's corotation radius is } $r_\mathrm{c}  \, {=} \,2 \, \mathrm{b}$
while the shape function, $r_\mathrm{s}(t,\varphi)$ {(resp. $r_\mathrm{l}(t,\varphi)$}), encode{s} the {geometrical aspect of the spiral (resp. lump).}\\

  \noindent For the spiral linked to the binary period we {set} $\Omega_\mathrm{s}  \, {=} \, \Omega_\mathrm{orb}   \, {=} \, \sqrt{ \mathrm{G}M/b^3}${, where the Keplerian approximation, as used e.g. by  \cite{lopez_armengol_circumbinary_2021}, is sufficiently accurate at the separations considered here and for the purpose of our simple model (a more accurate description can be obtained via the post-Newtonian equation of motion, e.g. \citealt{mignon-risse_impact_2023}).} 
  {The} shape function {reads}
 \begin{equation}
      r_\mathrm{s}(t,\varphi)  \, {=} \, r_\mathrm{c} \  e^{\left( \delta_{\mathrm{s},\varphi} (\varphi-\Omega_\mathrm{s} t)\right)}
   \end{equation}
   {with an azimuthal extent set to $\delta_{\mathrm{s},\varphi} \, {=} \, 0.1$, compatible with \cite{cimerman_gravitational_2023}; in the general case, it should depend on the disk aspect ratio with thinner, colder disks producing sharper spiral waves, but we only consider one value here to focus on other parameters (e.g. the perturbation's amplitude $\gamma_\mathrm{s/l}$).}

For the lump, {orbiting at frequency}  $\Omega_\mathrm{l}  \, {=} \, \sqrt{ \mathrm{G}M/ {r_\mathrm{c}}^{3}}$, 
  the shape function is
\begin{equation}
      r_\mathrm{l}(t,r,\varphi) = e^{\left( -(r-r_\mathrm{c})^2/ \delta_\mathrm{l,r} \right)} \  e^{\left( - (\varphi-\Omega_\mathrm{l} t)^2/  \delta_\mathrm{l,\varphi}  \right)}
\end{equation}
 {with radial and azimuthal extent parameters $ \delta_\mathrm{l,r}\, {=} \, 0.1$ and $\delta_\mathrm{l,\varphi} \, {=} \, 0.6$, respectively}\footnote{{This gives a radial extent within the range reported in \cite{noble_mass-ratio_2021} from the density distribution, and an azimuthal extent close to their lower-range value.}}.
    With those assumptions we {fixed} $\Omega_\mathrm{l} \, {=} \, \Omega_\mathrm{s}/5$ {which sets $r_\mathrm{c}$ slightly beyond} $r_{\mathrm{in}}$, 
    the edge of the CBD.
    These are approximations, as the inner edge of the CBD is not exactly at twice the separation and the cavity's eccentricity {(e.g. \citealt{duffell_santa_2024}), although lower in GR than in Newtonian dynamics (\citealt{noble_circumbinary_2012}, \citealt{noble_mass-ratio_2021})} is neglected here. 
    
  \begin{figure}
\includegraphics[width=0.48\columnwidth]{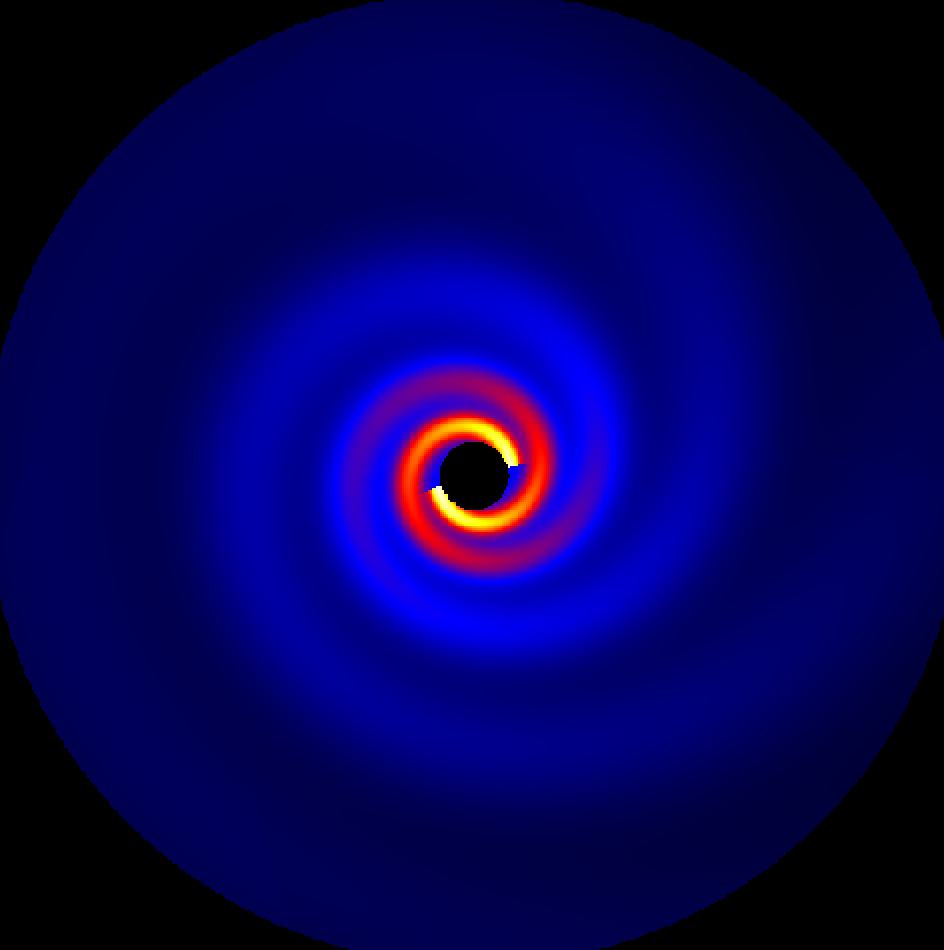}
\includegraphics[width=0.48\columnwidth,height=0.48\columnwidth]{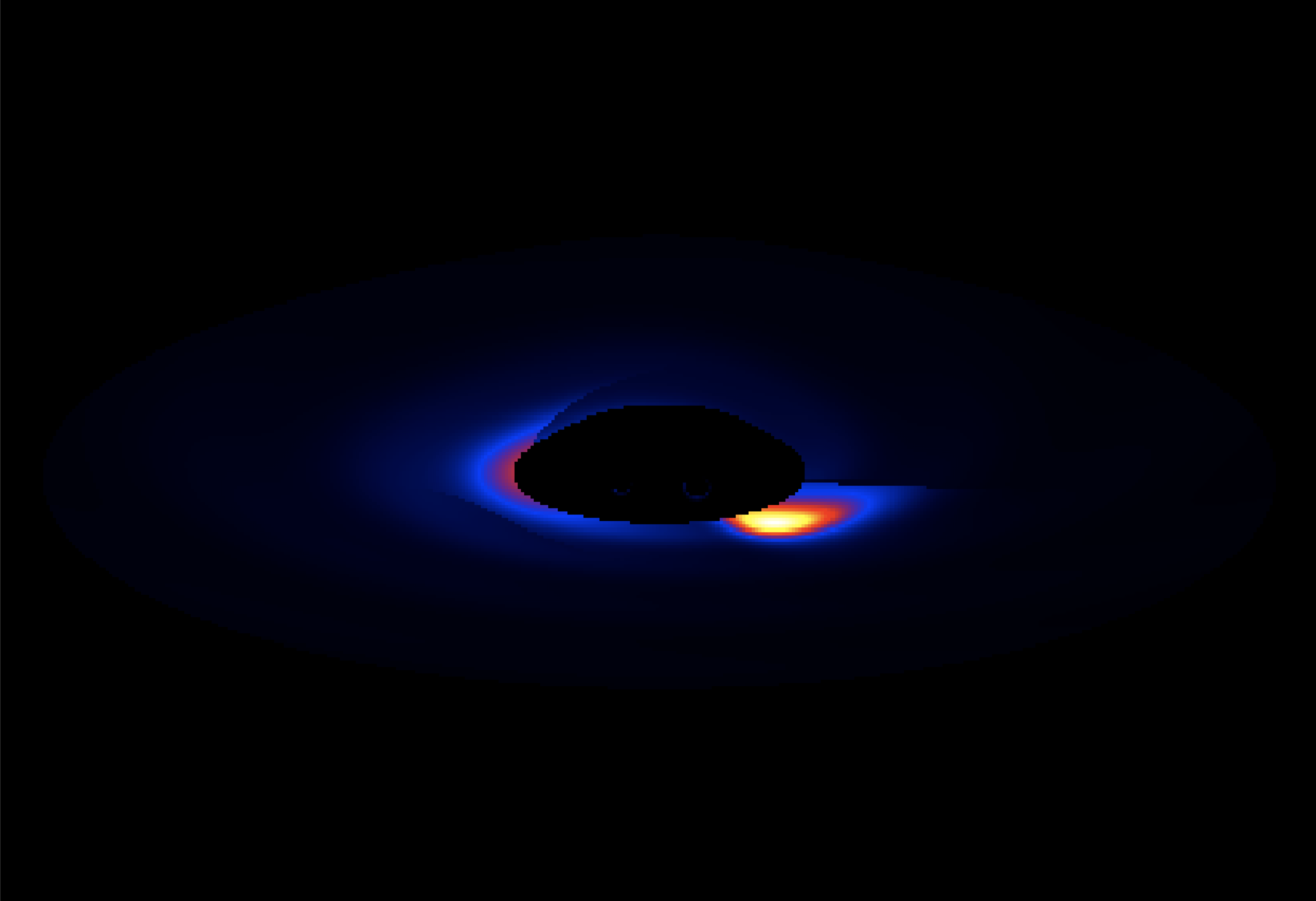}
\caption{Emission maps with the \{disk + spiral\} {temperature distribution} computed at inclination $i\, {=} \, 20^\circ$ (left panel; $\gamma_\mathrm{s}\, {=}\,0.5$) and \{disk + lump\} {temperature distribution} computed at inclination $i\, {=} \, 70^\circ$ (right panel; $\gamma_\mathrm{l} \, {=}\,0.3$).
{The value of $\gamma_\mathrm{s}$, higher than in the rest of the manuscript, has been chosen to enhance visibility.}
}
\label{fig:snap}
\end{figure}         
The impact of such non-axisymmetrical structure in the temperature profile has been previously shown to lead in single {BH disks} to quasi-periodic {electromagnetic} modulation \citep{varniere_rossby_2019}{, in particular due to the relativistic Doppler effect \citep{vincent_flux_2013}}. 

\subsection{Ray-tracing with {\tt gyoto}}
\label{sec:gyoto}

The previously presented temperature distribution is then used as an input for the GR ray-tracing code 
{\tt gyoto}\footnote{http://gyoto.obspm.fr} \citep{vincent_gyoto_2011} in a {time-dependent} 
BBH spacetime
{valid down to $b\, {\sim} \ 8$~M}
 (the so-called Near Zone, \citealt{mundim_approximate_2014}, \citealt{ireland_inspiralling_2016}, \citealt{mignon-risse_impact_2023}; 
including the Far Zone as well, valid beyond a gravitational wavelength from the binary, \citealt{johnson-mcdaniel_conformally_2009}) to obtain 
multi-wavelength thermal emission maps {as shown} in Fig.~\ref{fig:snap}, and from there, the LCs.
It might seem contradictory to couple our simplified disk model with a full ray-tracing {method}, but the importance of performing such ray-tracing was previously shown 
when studying variability in {single BH} disk{s} (\citealt{varniere_impact_2016}, \citealt{casse_rossby_2018}).

One example is the relativistic Doppler effect{, whose amplitude can be roughly estimated in an equivalent Schwarzschild metric (e.g. \citealt{tang_late_2018})}.
Indeed, the intensity amplification is proportional to $\mathcal{D}^3$ with $\mathcal{D}\, {=} \, [\Gamma (1- v_\mathrm{\parallel})]^{-1}$, where $v_\mathrm{\parallel}$ is the gas velocity parallel to the observer's line-of-sight defined as $v_\mathrm{\parallel}= v \cos(\phi) \sin(i)$, with $v$ the gas velocity, $\phi$ the azimuthal angle with respect to the line-of-sight and $i$ the inclination angle.
Considering an orbiting, warm lump as it is approaching the observer, we simply use $v_\phi$, the azimuthal velocity, as $v \cos(\phi)$ and, as it is orbiting {close to} the local Keplerian velocity, we get
\be
\mathcal{D}^3(r_\mathrm{l},i) \approx \left[ \frac{1 - (r_\mathrm{l}/M)^{-1} }{1 - (r_\mathrm{l}/M)^{-\frac{1}{2}} \mathrm{sin}(i) } \right]^3.
\label{eq:D3}
\ee
For  $r_\mathrm{l} \, {=} \, 2$~b, $b\, {=} \, 20$~M, and $i \, {=} \, 70^\circ$ this gives an intensity boost equal to $\mathcal{D}^3 \, {\approx}\,1.5$, justifying the need to incorporate these relativistic effects.

The CBD is assumed to be optically thick and fully located in the $z\, {=} \, 0$ plane.
The emission follow{s} Planck's law and the  intensity is computed as $I_\nu  \, {=} \, B_\nu(T(r,\varphi))$ with $B_\nu$ the Planck function.
Null geodesics are backwards integrated from the observer to the emitting plasma, until they hit the disk plane.
Hence, absorption and scattering are neglected here.  When the geodesics intersect the plasma, its {temperature and velocity} at the time of emission are taken from the model.
{In this procedure, the time-dependence of the metric is accounted for because photons travel with a finite velocity (no \textit{fast-light approximation}).}
\newline 

{Unless stated otherwise,} we will compute spectr{a} and LC{s} for a distance $D \, {=} \, 500$~Mpc, a binary total mass $M \, {=} \, 4\, {\times} \, 10^6 \, \mathrm{M_\odot}$ and an accretion rate $\dot{M} \, {=} \, 0.5 \, \mathrm{\dot{M}}_\mathrm{Edd}$.
{However, at fixed accretion rate in Eddington units, our temperature normalization assumes $ T \, {\varpropto} M^{-1/4}$ (Eq.~\ref{eq:T}), so the total flux integrated over a source of surface $ {\varpropto} \, r_\mathrm{g}^2\,  {\varpropto} \, M^2$ scales as $F \, {\varpropto} \, T^4 r_\mathrm{g}^2 \, {\varpropto} \, M$ (e.g. \citealt{roedig_observational_2014}).
Thus, a}s we focus on the {shape of the} LC modulation{, we will compute the frequency-integrated flux and renormalize it by the time-averaged flux: the resulting LCs, once normalized, are mass-independent.}

\section{An energy-dependent variability for the CBD}
\label{sec:LC}

From previous work, we know that the presence of non-axisymmetrical structures in the temperature profile {of BH disks} will lead to some almost periodic oscillations of the
disk {thermal radiative emission} \citep{varniere_rossby_2019}. 
{Here we study these in the context of CBDs, in an approximate BBH spacetime (instead, a Schwarzschild metric was used in \citealt{tang_late_2018}), including e.g. gravitational redshift, relativistic Doppler effect, and varying the temperature perturbations' parameters.
We will investigate their shape, energy dependence and the influence from intrinsic system's parameters.}

\subsection{Energy-dependence of the variability}
\label{sec:PDS}  

  One of the interesting observational aspect{s} of a{n orbiting} non-axisymmetrical structure in the temperature {distribution} resides in the changing shape of its {SED} along that structure's orbit \citep{varniere_living_2020}. 
  {As we focus here on the temporal evolution of the SED for our CBD model, we consider, as a first example, a} 
  separation  $b\, {=}\,20$~M, {and temperature perturbations with} $\gamma_\mathrm{l} \, {=} \, 0.{1}$ {($T/T_0{\approx}\, 20\%$)} and $\gamma_\mathrm{s} \, {=} \, 0.0{2}$ {($T/T_0\, {\approx}\, 4\%$)}. 
   \\ 

\begin{figure}
\includegraphics[width=\columnwidth]{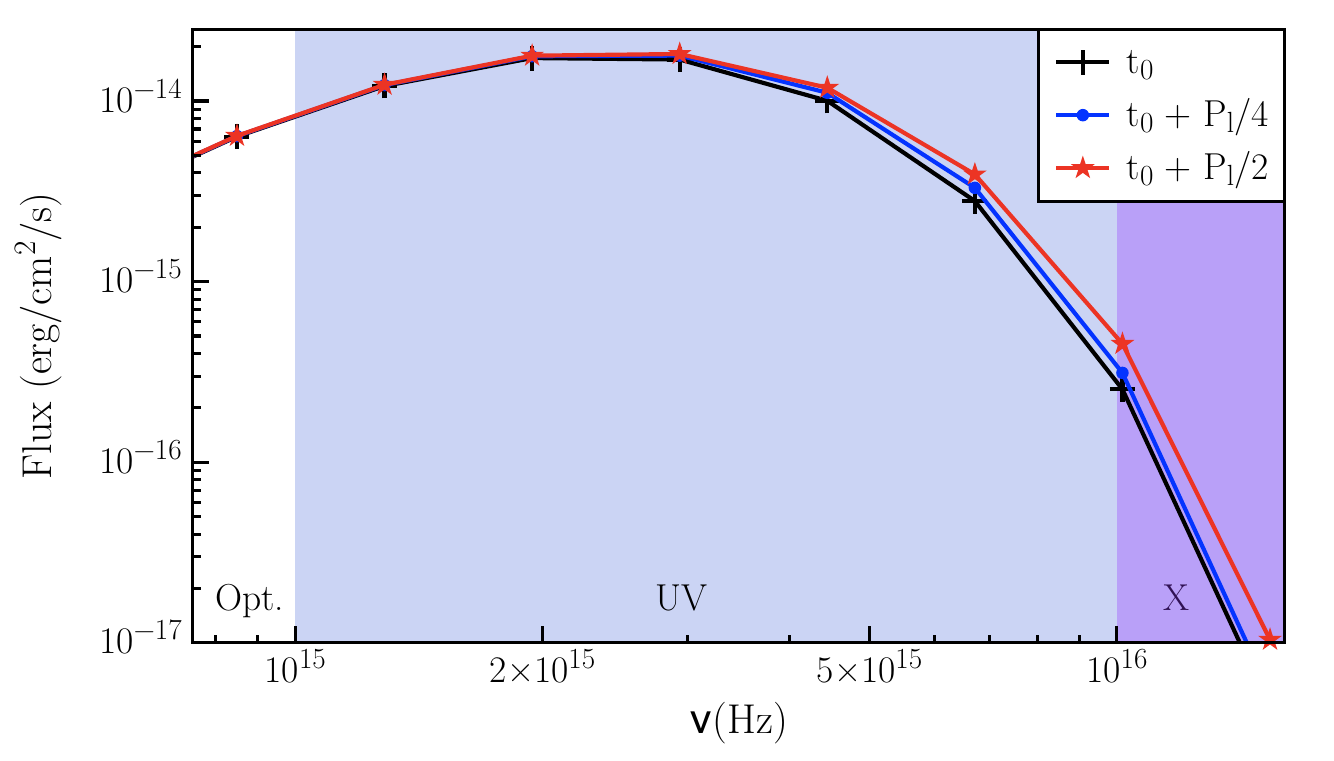}
\caption{Evolution of the spectral energy distribution along {the lump's orbit with $P_\mathrm{l} \, = \, 2\pi / \Omega_\mathrm{l}$}, for model parameters $\gamma_\mathrm{l}\, {=}\,0.{1}$, $\gamma_\mathrm{s}\, {=}\,0.0{2}$, $b\, {=}\,20$~M
for a mass  $M \, {=} \, 4\times 10^6 \mathrm{M_\odot}$ seen at an inclination angle of $70^\circ$.
{The initial time here, $\mathrm{t}_0$, is chosen to correspond to a minimal value of the flux.}
While the SED stays the same below $10^{15}$~Hz, it is pivoting at higher {frequency} as the lump moves along its orbit.}
\label{fig:SED}
\end{figure}

Figure~\ref{fig:SED} shows the energy spectrum, at several {instants along a portion of} the lump's orbit. We see that at {frequency} below $10^{15}$~Hz
for our choice of mass  $M \, {=} \, 4\, {\times} \, 10^6\,  \mathrm{M_\odot}$, the SED does not show any noticeable changes.
{Mean}while{,} at higher {frequency} the SED {pivots} toward higher flux {because} it is linked to the innermost, warmest parts of the CBD{; this behavior is thus, qualitatively, independent of the perturbations' parameters}.
{The flux increases by a factor ${\approx}1.1$ at $3\times10^{15}$~Hz, where the (mean) SED peaks, and ${\approx}2$ at $10^{16}$~Hz}. 
{Thus, as a function of time}, it will create a strong modulation of the LC along the lump's orbit.
{The amplitude of the flux modulation} will {naturally} depend on the frequency band at which it is computed.
\\

{
This case illustrates, via the temporal evolution of the SED, the frequency-dependence of the expected flux modulation.
However, a}s we want to present mass-independent LCs, in the following we will present LC{s computed by integrating the spectrum over all frequencies and normalized to the time-averaged bolometric flux (as mentioned in Sec.~\ref{sec:gyoto}).} 

\subsection{Impact of the binary parameters on the variability}

\begin{figure}
\includegraphics[width=\columnwidth]{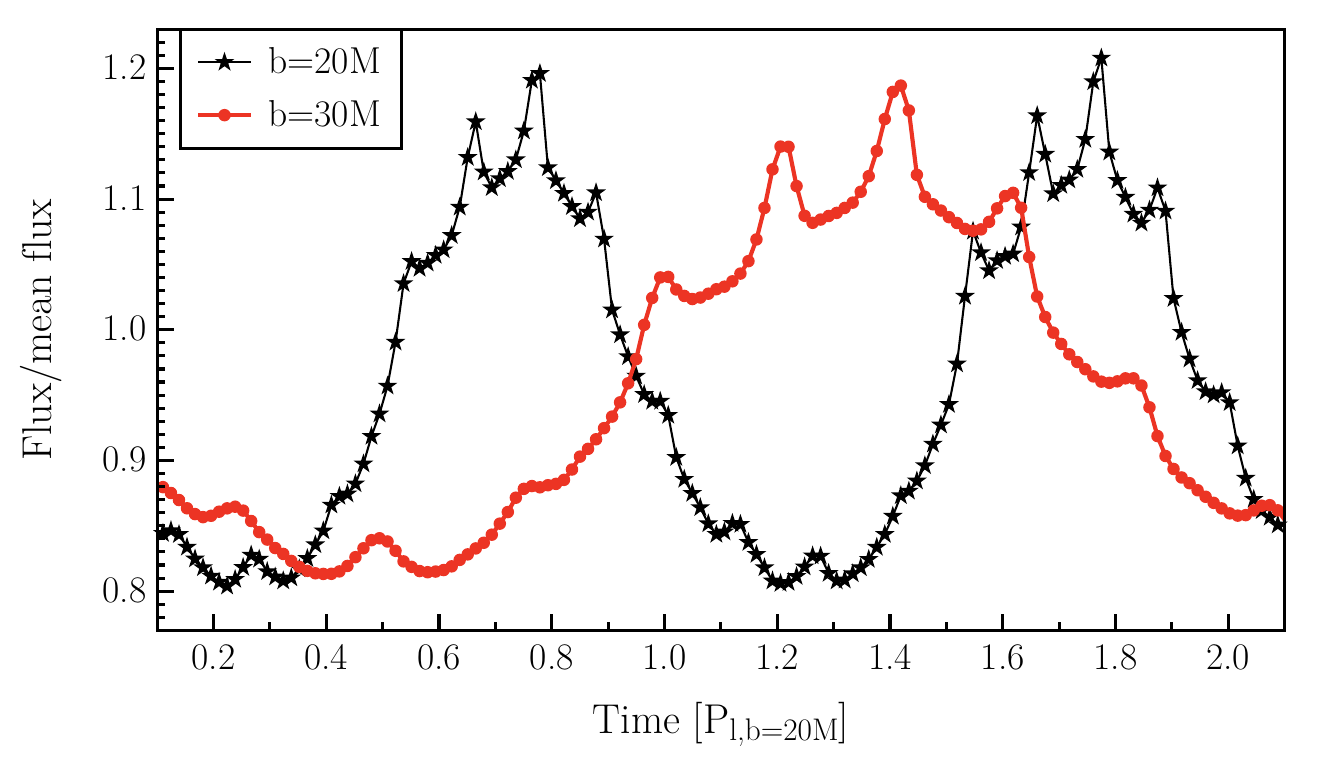}
\caption{Bolometric flux, normalized {by the average between the minimal and maximal values}, as a function of time, for $b\, {=}\,20$~M (black curve and stars) and $b\, {=}\,30$~M (red curve and dots). Both LCs are computed for 
$\gamma_\mathrm{l}  {=}\,0.3$, $\gamma_\mathrm{s}\, {=}\,0.05$, $i\,{=}\,70^\circ$, with $0^\circ$ corresponding to face-on view.}
\label{fig:2b}
\end{figure}
One benefit of using a simple parametric model is that we can explore how the binary or orbital parameters influence the resulting flux modulation {at reasonable cost.
Beyond the expectation of a modulation, non-trivial outcomes are the modulation's shape and amplitude, in particular in a BBH metric, and their dependence on the system's parameters}.
As the inner edge {radius} of the CBD and the binary period are only slightly dependent on the mass ratio \citep{noble_mass-ratio_2021}, {a more} important binary{-related} parameter {to be varied} is the separation (see {S}ec.~\ref{sec:repr_cbd}), especially {in the context of} mass-independent LCs.

  From Fig.~\ref{fig:SED} we expect a modulation of the flux linked with the lump's orbital period, {which is itself linked to} the binary {period and therefore its} separation {(Sec.~\ref{sec:repr_cbd}).}
  {If similar to the single-BH case} \citep{varniere_impact_2016}, we {would} also
  expect a smaller modulation coming from the spiral orbiting at half the binary period. 
 This is confirmed by the LCs of Fig.~\ref{fig:2b} which shows, at two different orbital separations ($b\, {=}\,20$~M and $b\, {=}\,30$~M) the normalized, bolometric LC  
  of the same {parameters for the temperature perturbations} ($\gamma_\mathrm{l} {=}\,0.3$ and $\gamma_\mathrm{s}\, {=}\,0.05${, higher values than previously to enhance visibility of the smaller-amplitude modulation}).

In both cases, two modulations are visible on the LC, with the dominant one being of the order of ${\approx}\,20\%$, while the {other one, with a shorter} period{,} reach{es} a maximum of ${\approx}\,5\%$.
{The dominant modulation exhibits a roughly sinusoidal shape}.
As expected, the dominant modulation comes from the orbiting lump while the smaller amplitude one is due to the two-arms spiral propagating in the disk at the binary period. 
\newline

\subsection{Impact of {the amplitude of the temperature perturbation}}

\begin{figure}
\includegraphics[width=\columnwidth]{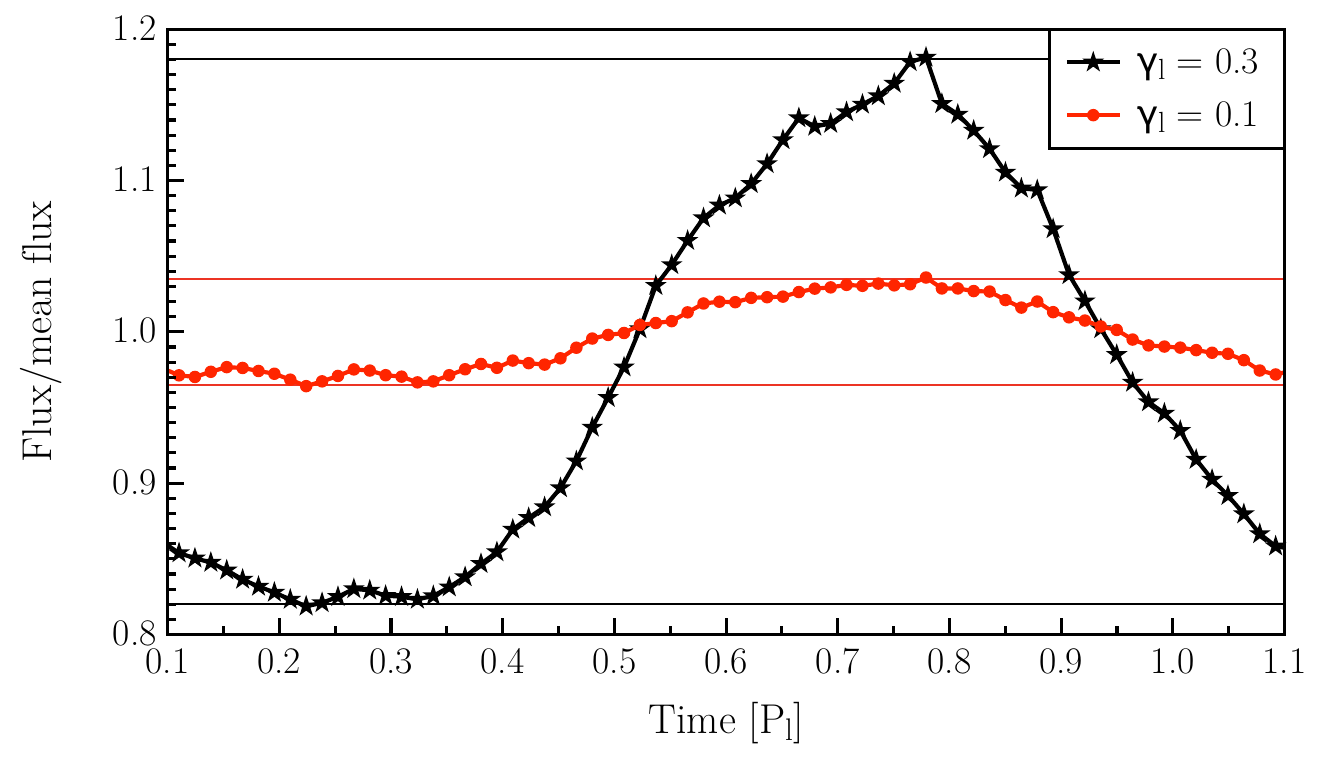}
\caption{Bolometric flux, normalized {by the average between the minimal and maximal values}, as a function of time, for $\gamma_\mathrm{l} \, {=}\,0.3$ (black curve and {stars}) and $\gamma_\mathrm{l} \, {=}\,0.1$ (red curve and {dots}{, corresponding to Fig.~\ref{fig:SED}}). 
The other parameters are fixed to $i\,{=}\,70^\circ$, $\gamma_\mathrm{s}\, {=}\,0.02$, $b\, {=}\,20$~M.}
\label{fig:diffgammal}
\end{figure}

   One of the controversial aspect{s} of the \lq lump\rq\ {-- a density feature by definition \citep{shi_three-dimensional_2012} --} orbiting at the edge of the {CBD} is by how much, if any, will it {convert into} an overheated 
   lump {as well} in the temperature {map} \citep[see for example][for different heating and cooling processes and their consequences on the lump]{noble_circumbinary_2012,farris_characteristic_2015,wang_role_2023}.
   While there is some consensus about a modulation of the flux at near the period of the lump (e.g. \citealt{tang_late_2018}, \citealt{cocchiararo_electromagnetic_2024}), the details about its origin are still debated.
   On top of that, the different simulations of CBDs, with various heating and cooling processes, rarely express quantitatively how the overdense lump structure translates 
   into the temperature distribution which renders studying the impact of the different heating efficiency difficult. 
   By using our model, where the heating efficiency is encapsulated in one parameter, {the temperature perturbation's amplitude,} we can explore how a less efficient  heating of the lump would affect the 
   LC and therefore the detectability of the modulation. 
\newline
  
   In Fig.~\ref{fig:diffgammal} we compare the amplitude of the \lq main\rq\ lump modulation when the 
   {associated temperature perturbation has $\gamma_\mathrm{l} \, {=} \, 0.3$,}   
    versus the more conservative value $\gamma_\mathrm{l} \, {=} \, 0.1${, both for $\gamma_\mathrm{s} \, {=} \, 0.02$}. 
    We see that in both cases the lump {remains the origin of} the strongest modulation. 
       {M}ore quantitatively we see that $\gamma_\mathrm{l}  \, {=} \, 0.3$ lead{s} to a flux modulation {of $\pm 18\%$ amplitude,}
   while reducing {it} by a factor of three gives a flux modulation {of around $\pm4\%$ amplitude.
   In the latter case, although the spiral-related modulation is present by construction, it is no longer visible by eye.}
   
   Such trend between the {temperature perturbation's amplitude} and the amplitude of the flux modulation implies that even a few percent {temperature perturbation} could lead to 
   a small, but measurable, modulation of the flux.
   {Indeed, such an amplitude was seen in the optical band, where the BBH candidate PG 1302-102 exhibits a modulation of $0.14$~mag \citep{graham_possible_2015}, i.e. ${\approx}13\%$ in our units.
   Naturally, the lower the amplitude, the higher the number of periodic cycles covered by the data is needed for a convincing detection (see e.g. \citealt{cocchiararo_electromagnetic_2024}).
   }

\section{Observational toy model for the CBD variability}
\label{sec:toy}

    Now that we have {illustrated} how the LC and its modulations relate {to the main BBH intrinsic} parameters {for a few cases}
    , we 
    {focus on source-related parameters} 
    such as the impact of the
    distance, inclination on the observables.
    {We} also {study} how the total mass of {the} binary impact{s the modulation's timescale and optimal frequency band}.
    From all of this we will then present a simple toy model that will 
    {help the} search for CBD variabilities.

\subsection{Impact of the distance and inclination}
    
    {The observed radiative} flux {of a BBH system} depend{s} on {the source's} distance {${D}$} and inclination {angle $i$}. 
    For the 
    multicolor blackbody 
    emission\footnote{see the diskbb model at \url{https://heasarc.gsfc.nasa.gov/xanadu/xspec/manual/node165.html}} used here,
    both apply to the flux normalization which is   $(r_\mathrm{in}/D)^2 \cos i$. 
        
    Things are more complicated when we look at the inclination as it is also related to GR effects. 
    {T}hose GR effects 
    tend to increase the amplitude of modulation{s} coming from an orbiting non-axisymmetric structure
    through Doppler boosting, with an increasing impact as the inclination increases (from face-on to edge-on) and as the separation decreases, as shown in Eq.~\ref{eq:D3}.
{For completeness, we note }that, at the distance {separating the CBD} from the BBH, the gravitational beaming is subdominant.

     \begin{figure}
\includegraphics[width=\columnwidth]{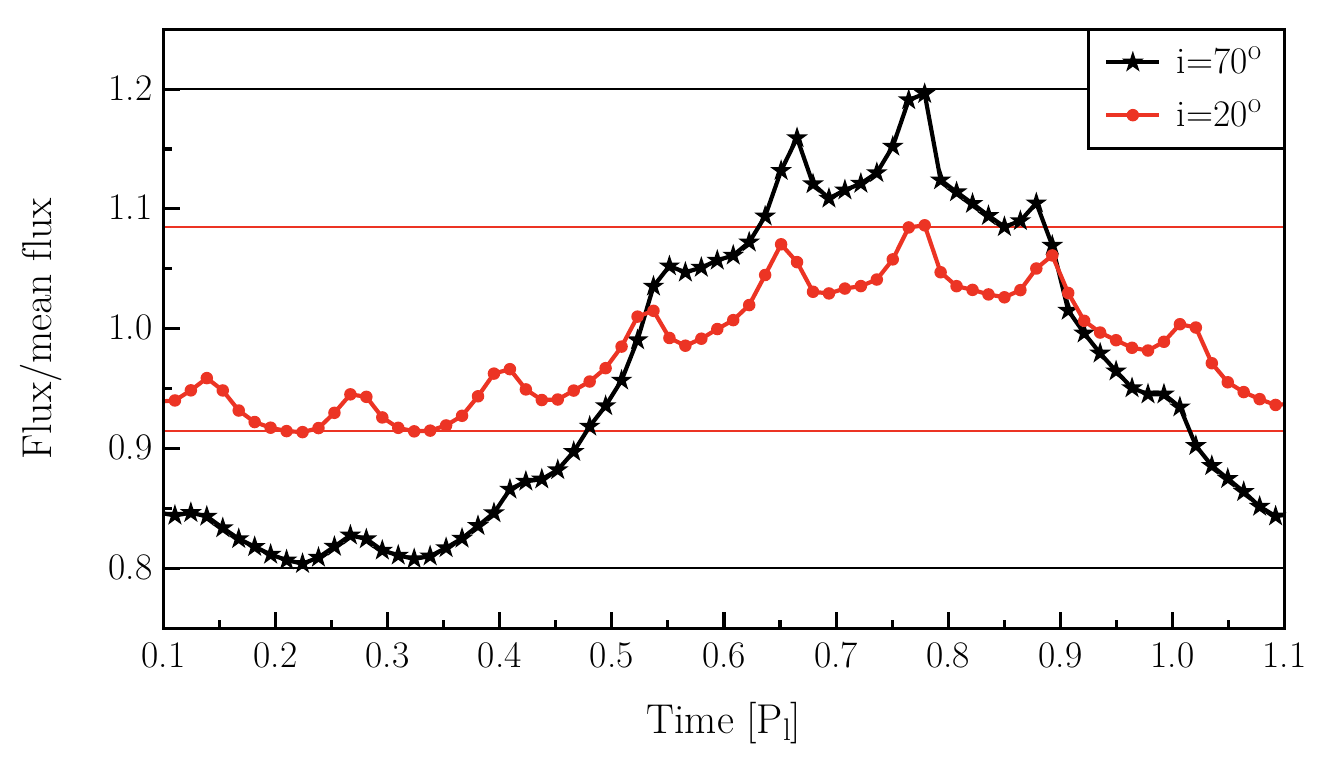}
\caption{Bolometric flux, normalized  {by the average between the minimal and maximal values}, as a function of time, for $i\,{=}\,70^\circ$ (black curve and stars) and $i\,{=}\,20^\circ$ (red curve and dots). The other parameters are fixed to $\gamma_\mathrm{l}\, {=}\,0.3$, $\gamma_\mathrm{s}\, {=}\,0.05$, $b\, {=}\,20$~M.}
\label{fig:diffincli}
\end{figure}
    
     This is showcased in Fig.~\ref{fig:diffincli} which compares the amplitudes of the observed modulation from the same disk seen at two distinct inclinations,
     one more edge-on at $70^\circ$ and one closer to face-on at $20^\circ$. 
     While the modulation is still present and detectable, the low inclination LC has an amplitude of $\pm 10\%$, compared with $\pm 24\%$ at higher inclination.
     This means that {modulations from} systems seen close to face-on will be harder to detect. 
    \newline

\subsection{Mass dependence and optimal observation band}

\begin{figure}
\includegraphics[width=\columnwidth]{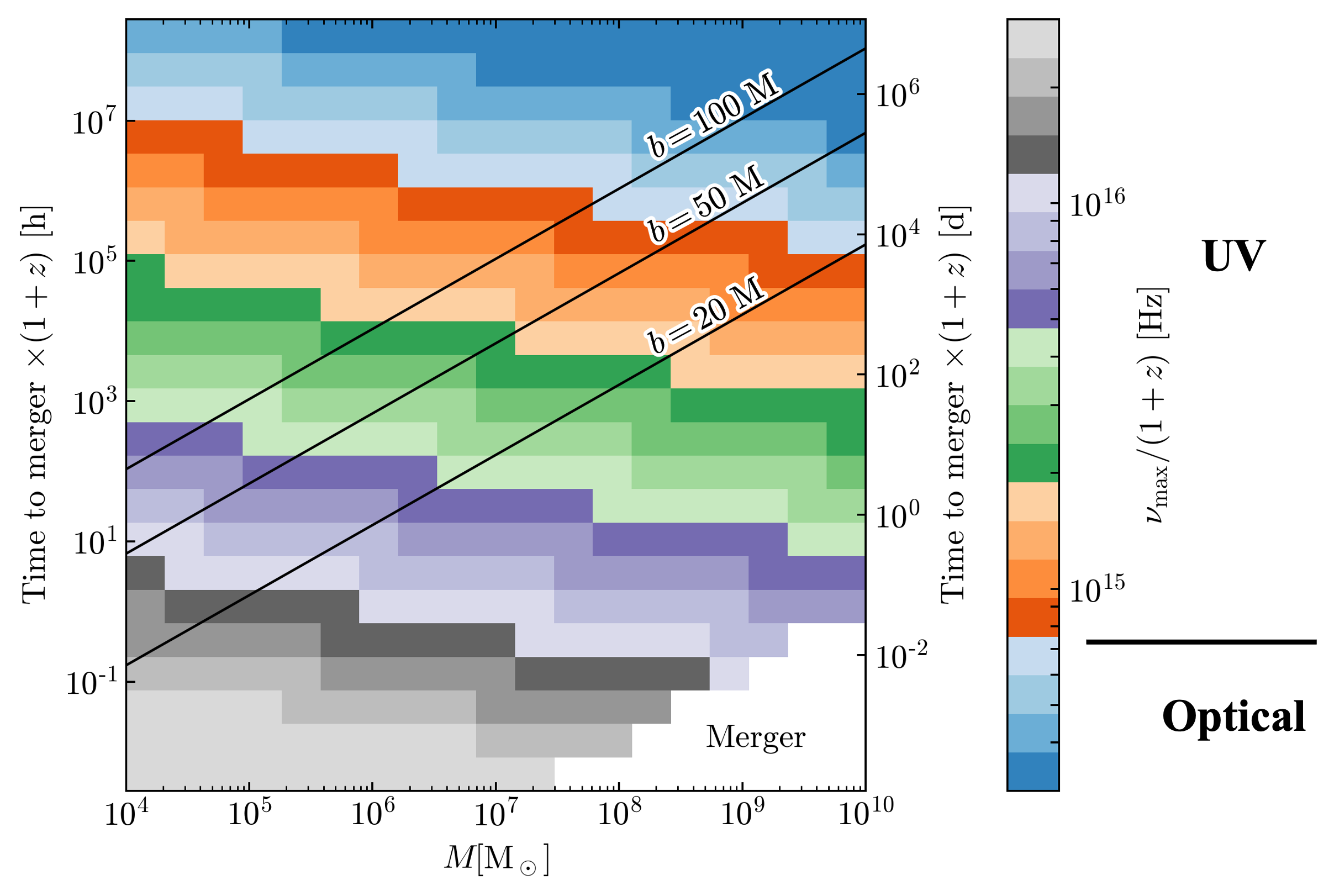}
\caption{Colormap of the emission spectrum peak frequency of the CBD, $\nu_\mathrm{max}$ (in Hertz) as a function of the total binary mass $M$ and the time to merger (in hours on the left, in days on the right) 
 derived from Eq.~\ref{eq:T}.
This approximates the optimal energy band to observe the LC modulations.
The "merger" region is defined as $b\le1$~M for simplicity.}
\label{fig:m_tmerger_numax}
\end{figure}

{In addition to} the distance and inclination, {an}other variable related to the source is its mass. 
Until now we have shown {renormalized} LC{s}, hence removing the{ir} mass-dependence.
{As already suggested by the SED shown} in Fig.~\ref{fig:SED}, {computing the LC by integrating the flux over all frequencies or
only in the frequency range } around the emission peak {give similar results}.

In order to find the optimal observation {frequency} band for the LC modulation, we need to estimate, as function of the binary mass and separation, the frequency at which 
the {SED} of the CBD peaks{, denoted ${\nu_\mathrm{max}}$.}
We obtain it by applying Wien's law to the CBD inner edge temperature (see Eq.~\ref{eq:T}).
We show in Fig.~\ref{fig:m_tmerger_numax} the colormap of ${\nu_\mathrm{max}}$ as a function of $M$ and the time to merger\footnote{{Here given for $q\, {=} \, 1$. }To obtain the time to merger for other values of $q$, it should be multiplied by $(1+q)^2/4q$.}. 
The time to merger has been evaluated using the leading-order expression \citep{peters_gravitational_1964} {; it gives the correct order-of-magnitude estime but becomes less accurate closer to merger.} 
Most notably, Fig.~\ref{fig:m_tmerger_numax} shows that the CBD peak frequency $\nu_\mathrm{max}$ is in the UV band for most {SM}BBH system parameters{, in agreement with most simulations (e.g. \citealt{cocchiararo_electromagnetic_2024}, \citealt{tiwari_radiation_2025}).}
For {PTA sources} ($M\, {\ge} \, 10^8 \mathrm{M_\odot}$), $\nu_\mathrm{max}$ is in the optical band when the orbital separation is large enough, corresponding to decades away from merger. 
{This coincides with the band hosting most variable sources attributed to BBH candidates \citep{graham_systematic_2015}}.

\subsection{Toy model reproducing the main EM variability of the CBD}

{T}he results from the previous sections {suggest} that it is possible to create a simple analytical toy model able to reproduce {the }
 electromagnetic {variability} of {the CBD model}.
 Indeed, {we will show that} the previous {renormalized} LCs {can} be fitted by a simple sum of {sinusoidal functions} on top of the axisymmetric disk emission. 
This leads to the following model:
\be
F(t)=F_0 \left[1 + A \cos \left( \frac{2 \pi t}{\mathrm{P_{l}}} + \phi \right) + A' \cos \left(  \frac{2 \pi t}{\mathrm{P'}} + \phi' \right) \right] 
\label{eq:model},
\ee
with  $F_0$ the unperturbed disk flux (with peak emission in the range shown on Fig.~\ref{fig:m_tmerger_numax}), $A$ and $A'$ the respective amplitude{s} 
of the lump and spiral-related modulations, $\phi,\phi'$ their respective phases\footnote{The amplitude of the modulation is related to
the disk {properties and to the distance to} the source. It is
therefore kept as a free parameter of the model. Similarly, the phase depend{s} on the {azimuthal} position of each structure at the start of the observation.
}.
{
$\mathrm{P_{l}}$ is again the lump period and $\mathrm{P'}$ the orbital-related period which results in $\mathrm{P_{orb}}/2$ for our symmetric spiral; for unequal-mass binaries, the spirals are likely to be asymmetric (e.g. \citealt{cimerman_gravitational_2023}) so the dominant one would produce a modulation at $\mathrm{P'}\, {=}\, \mathrm{P_{orb}}$ }.
{W}e can estimate the period of each modulation {as}
\begin{align}
\mathrm{ \frac{P_{orb}}{2}}  \ \  &{\approx} \ \ 1.4 \,  \ \  \left( \frac{b}{20\, \mathrm{M}}  \right)^{3/2} \,  \left( \frac{M}{10^6\, \mathrm{M_\odot}}\right) ~\ \ \mathrm{ks}     \\
 \mathrm{P_{l}}  \ \  &{\approx}  \ \   17  \ \ \   \left( \frac{b}{20\, \mathrm{M}}  \right)^{3/2}  \,\left( \frac{M}{10^6\, \mathrm{M_\odot}}\right) ~\ \ \mathrm{ks}  \ \ \ \ 
  \label{eq:periods}
\end{align}
with the lump period taken as an average between various numerical studies (e.g. \citealt{noble_circumbinary_2012}, \citealt{westernacher-schneider_multiband_2022}).
\begin{figure}
\includegraphics[width=\columnwidth]{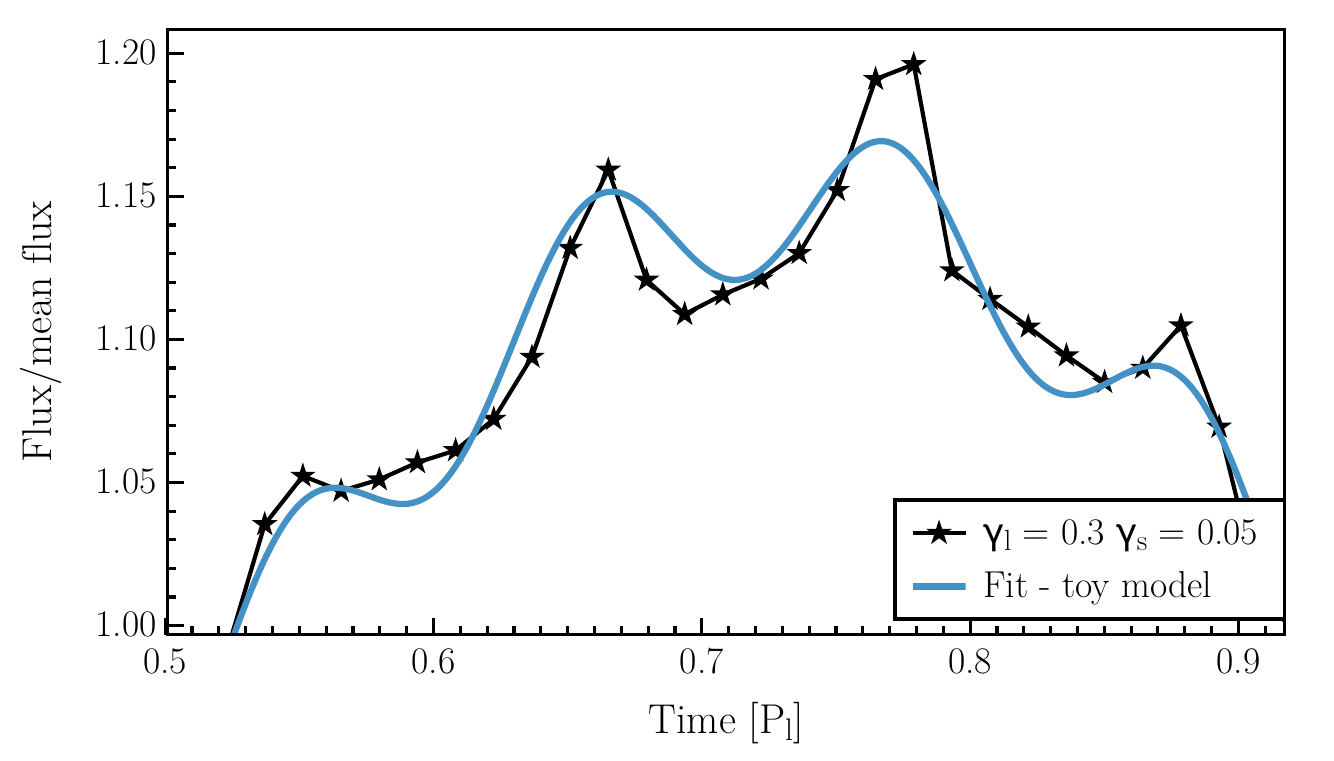}
\includegraphics[width=\columnwidth]{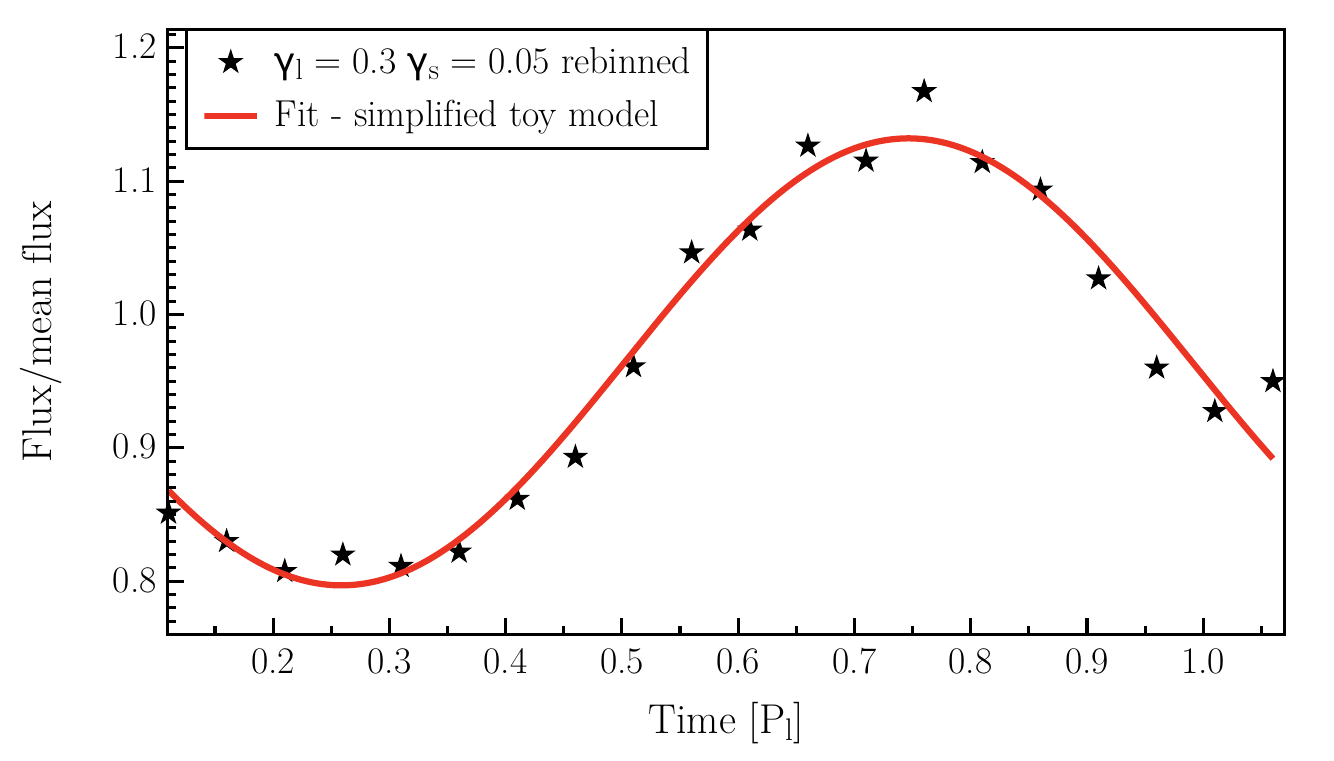}
\caption{
{
Top panel: bolometric flux from our temperature model, normalized by the average between the minimal and maximal values, as a function of time (black curve and stars), and its fit by the toy model function (Eq.~\ref{eq:model}, blue curve).
Bottom panel: rebinned, normalized lightcurve (stars) and its fit by the simplified toy model function (Eq.~\ref{eq:simplifiedmodel}, red curve).
The model parameters are fixed to $\gamma_\mathrm{l}\, {=}\,0.3$, $\gamma_\mathrm{s}\, {=}\,0.05$, $b\, {=}\,20$~M, $i\,{=}\,70^\circ$}.}
\label{fig:LCfit}
\end{figure}
{The $q$-dependence of $\mathrm{P_{orb}}$ has not been included as {this dependence} affects $\mathrm{P_{orb}}$ at the percent level only.
Similarly, the dependence of the lump orbital period on $q$, which corresponds to the Keplerian period at the inner edge of the CBD and thus depends on the CBD's inner radius, is smaller than its empirical uncertainty among numerical studies (Sec.~\ref{sec:key}). 
\\

How such toy model {(Eq.~\ref{eq:model})} would fit the LC obtained previously is shown on the top of Fig.~\ref{fig:LCfit}. We see that, while not being a perfect fit, it recovers the overall behavior of the LC, with residuals ${\sim}1\%$.
Such simplified model does not have the aim to give the exact LC of a CBD but is more geared toward producing a fast estimate of the potential variability and its
optimal observation range. \\

This model can be simplified even further, as  the modulation linked to the binary is not only much faster than the one from the lump but also has a much smaller amplitude{: thus, in practice, the spiral-related modulation might not be detectable}\footnote{{The mini-disc variability due to Doppler boost, not included here, occurs on the {same period as the spiral arms} (e.g. \citealt{dorazio_relativistic_2015}). The toy model might be extended to include it and the sum of both contributions might be detectable.}}.
As a result{,} we can omit its contribution (the third term in Eq.~\ref{eq:model}), especially if such period cannot be resolved by the available observations {(see our discussion in Sec.~\ref{sec:discu})}.
This gives the simplified version:
{
\begin{align}
F(t) =  F_0 \left[1 + A \cos \left( \frac{2 \pi t}{\mathrm{P_{l}}} + \phi \right)  \right]  \quad \text{if} \,  \Delta t_\mathrm{obs} \,{\gtrsim} \,  \mathrm{P'}/5,
\label{eq:simplifiedmodel}
\end{align}
}
\noindent where we assumed at least $5$ points per period to properly sample a given sinusoidal signal.  
{T}hat simpler version can always be used first and, by checking for any periodic structure in the residual, it can be decided if the full version of the model is needed.
The bottom panel of Fig.~\ref{fig:LCfit} shows the validation of the simplified toy model {(Eq.~\ref{eq:simplifiedmodel})} against a rebinned LC.
{By re-binning, we mean that we defined coarser time bins $\Delta t_\mathrm{obs}$ such that $\Delta t_\mathrm{obs} \,{>} \,  \mathrm{P_{orb}}/10$ and computed a flux average over each of these bins.}

\section{Comparison between the toy model and fluid simulations}
\label{sec:lc}

As a \lq test\rq\  of how this simple toy model {reproduces the main characteristics of more computationally-expensive} simulations, we  present here {various} fit{s} the 
LC{s} produced out of the $q \, {=} \,1$, $b\, {=} \, 20$~M {and $q \, {=} \, 0.1$, $b\, {=} \, 20$~M} run{s} presented in \cite{mignon-risse_origin_2023}{, together with an additional simulation with $q \, {=} \, 0.3$, $b\, {=} \, 36$~M} .

\subsection{Outlook of the simulation setup and post-processing}
\label{sec:simus}

In a nutshell, those {\tt GR-AMRVAC} \citep{casse_impact_2017} simulations consist in solving the conservation of baryon density, momentum in the same BBH background metric as used for the ray-tracing (Sec.~\ref{sec:gyoto}).
{The gas thermodynamical behavior is isentropic, so the total energy is conserved, and pressure evolves with the dynamics, unlike in the locally isothermal approximation.
A polytropic equation of state with adiabatic index $\gamma \, {=} \, 5/3$ is used.}
Initial conditions correspond to an axisymmetric, warm (aspect ratio $H/R\, {\sim}\,0.1$) disk close to centrifugal equilibrium around an equivalent ($M\,{=}\,M_1+M_2$) single BH.
After an initial transient phase, where the cavity self-consistently develops, spiral arms form in the CBD and the lump forms after tens of {BBH periods}.
Those remain present until the end of the simulation, i.e. after $100$ {BBH} periods.
For more details, we refer the reader to \cite{mignon-risse_origin_2023}.

\begin{figure}
\centering
\includegraphics[width=\columnwidth]{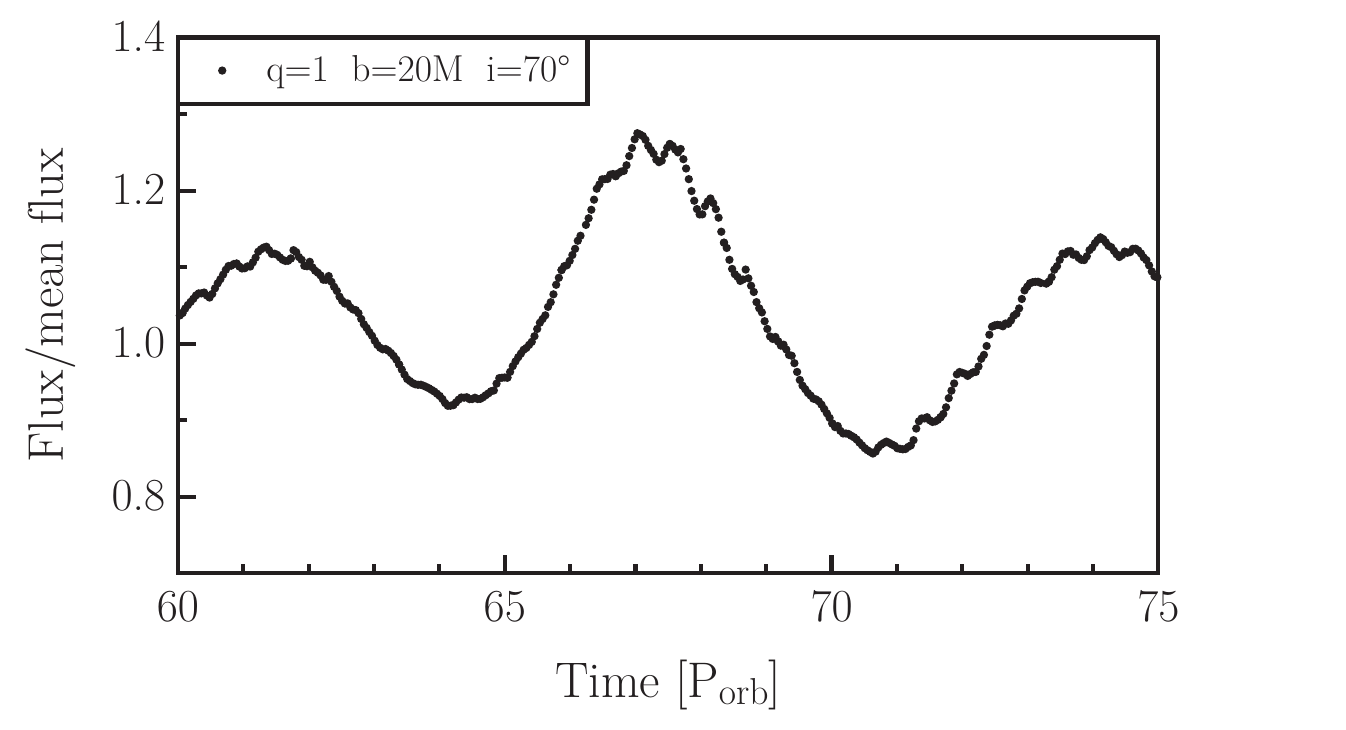}
\includegraphics[width=\columnwidth]{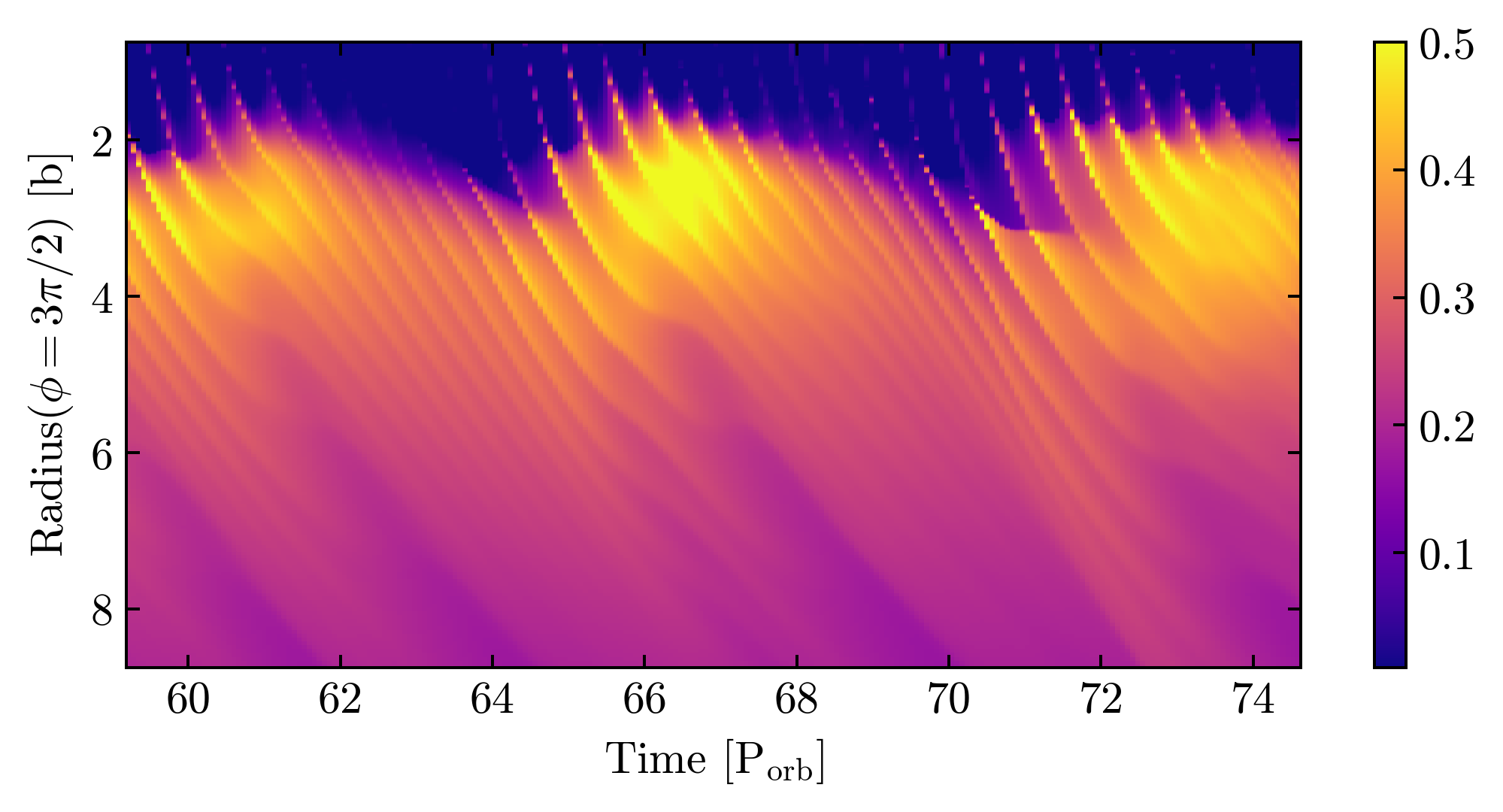}
\includegraphics[width=\columnwidth]{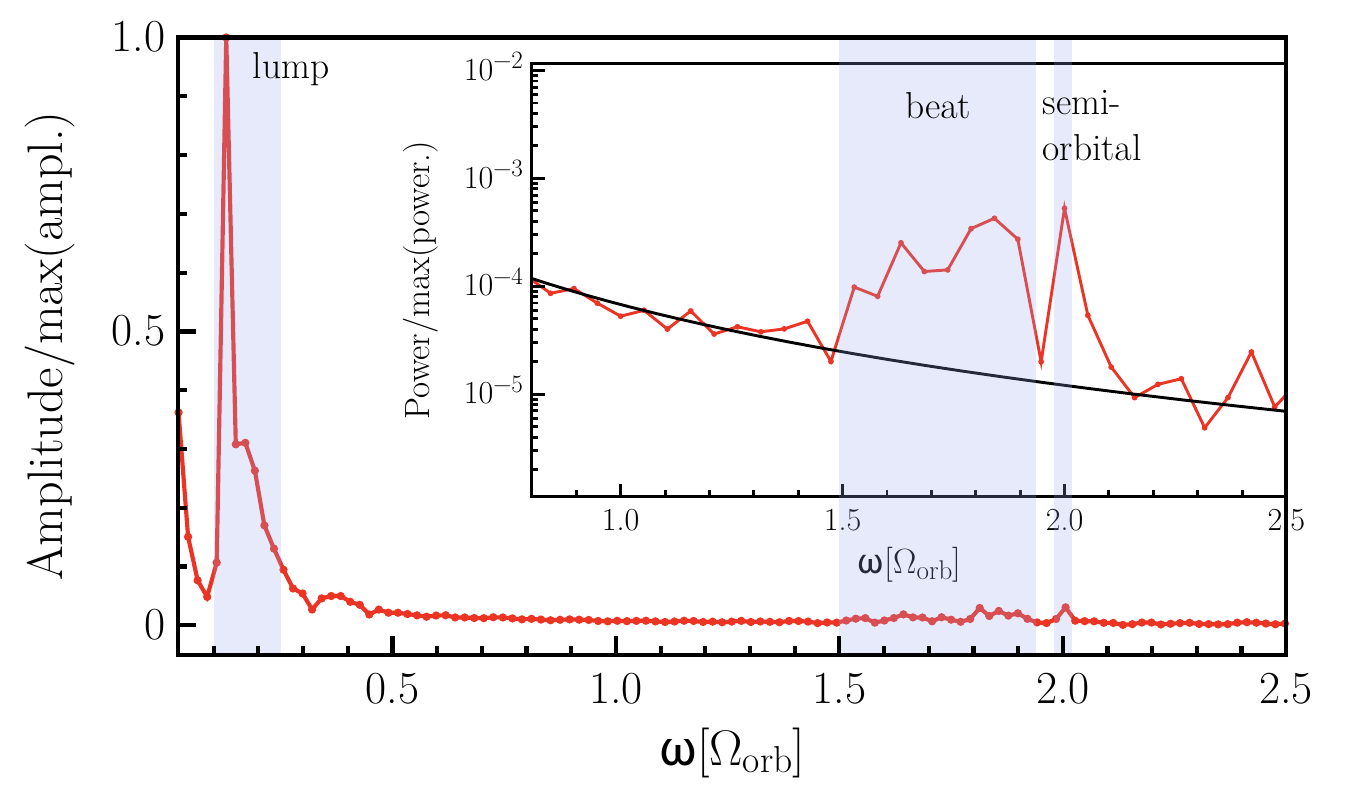}
\caption{Top panel: {normalized lightcurve} for $q\, {=} \, 1$, $b\, {=} \, 20$~M, seen an inclination angle $i \, {=} \, 70^\circ$.
Two modulations dominate the LC, at the semi-orbital period and at the lump period.
{Middle panel: density map in the $(t,r)$-plane for $\varphi \, {=} \, 3\pi/2$ ($\varphi \, {=}0$ being the line-of-sight, it maximizes the Doppler effect) over the same period}.
Bottom panel: amplitude of the Fourier modes{ in the LC}, dominated by the lump 
modulation{ and zoom-in view on the binary-lump beat signal, divided over a frequency interval, and the semi-orbital one.
Those are shown in power for enhanced visibility, and compared to a background red noise fit of the form $\omega^{-\alpha}$ with $\alpha\, {\approx}\, 2$}.}
\label{fig:LCq1}
\end{figure}

The fluid temperature {is computed} as $T\equiv P/ \Sigma$ (with $P$ the thermal pressure and $\Sigma$ the surface density) {which, in our model, comes down to considering only compressional heating and expansional cooling.}
{It is} used as an input for the ray-tracing step.
{Without the fast-light approximation, o}ne emission map calculation requires several {\tt GR-AMRVAC} outputs.
Since the time of emission does not, in general, coincide with an output time, the variable is linearly interpolated among the {\tt GR-AMRVAC} outputs.} \\

\subsection{Electromagnetic variability}
\label{sec:EM}

The top panel of Fig.~\ref{fig:LCq1} shows the bolometric LC normalized by the value averaged over the displayed time period ($15\, \mathrm{P_{orb}}$), obtained from the post-processed fluid simulation {with $q\, {=} \, 1$, $b\, {=} \, 20$~M, observed at inclination $i \, {=} \, 70^\circ$}.
Two modulations are visible, with the dominant modulation at a period equal to the lump orbital period ${\sim}6\, \mathrm{P_{orb}}$.
The orbital motion of the lump is illustrated by the fluid density map shown in the middle panel of Fig.~\ref{fig:LCq1} at the azimuthal angle $\varphi \, {=} \, 3\pi/2$.
{We obtain a non-axisymmetric temperature distribution.
This agrees with more sophisticated thermodynamical models, e.g. \cite{tang_late_2018} who include viscous heating, shock heating and radiative cooling.}
While this variability is persistent throughout the simulation (more than $60\, \mathrm{P_{orb}}$), {each modulation's period and amplitude} depend on the instantaneous local condition in the disk (e.g. \citealt{varniere_living_2020}). 
A Fourier analysis of the LC, as shown in the bottom panel of Fig.~\ref{fig:LCq1}, confirms that the lump modulation strongly dominates in amplitude.
{The peak at the lump frequency has a high statistical significance as it is about $20$ times the amplitude level at nearby frequencies.
The zoom-in view shows weaker features, at twice the binary-lump beat frequency (e.g. \citealt{roedig_limiting_2011} in hydrodynamics, \citealt{noble_circumbinary_2012}, \citealt{shi_three-dimensional_2012}, \citealt{combi_minidisk_2022}, \citealt{tiwari_radiation_2025} in magneto-hydrodynamics) distributed over a range of frequencies near $1.5\, \Omega_\mathrm{orb}$, and the spiral arms at frequency $2\,  \Omega_\mathrm{orb}$.
Despite an amplitude of about $6\, {-} \, 10$ times that of nearby frequencies, their firm detection is impeded by their weakness, the spread of the beat frequency and the proximity between both modes.
The potential presence of the beat frequency is interesting because it is intrinsically related to the hydrodynamical problem.
The amplitude of the orbital and beat modes could be amplified by the modulations from individual accretion structures, as seen in \cite{dorazio_relativistic_2015} and \cite{bowen_quasi-periodic_2018}, respectively.
}

\subsection{Toy model comparison and extension}
\label{sec:toy_comparison}

\begin{figure}
\includegraphics[width=\columnwidth]{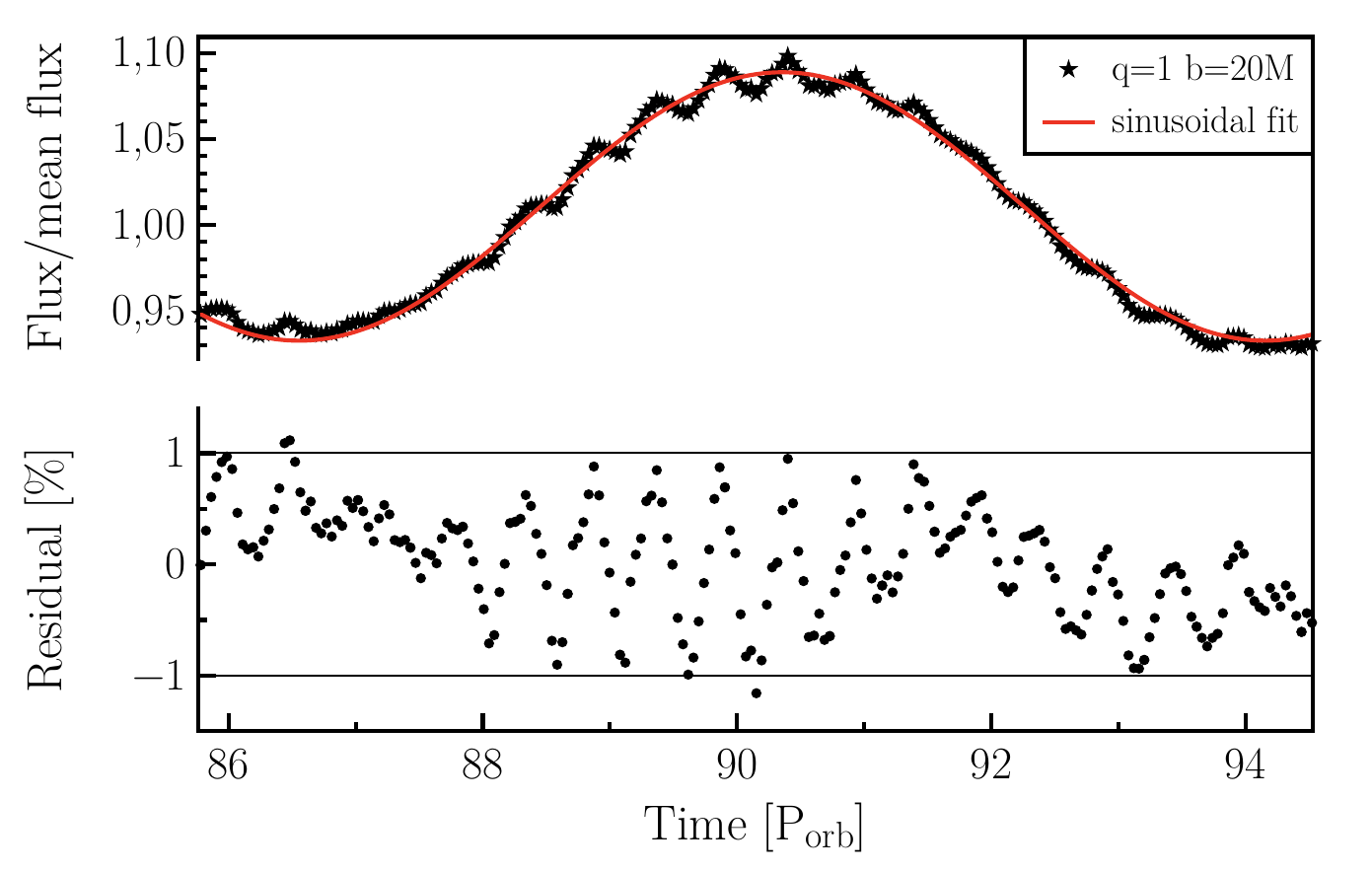}
\includegraphics[width=\columnwidth]{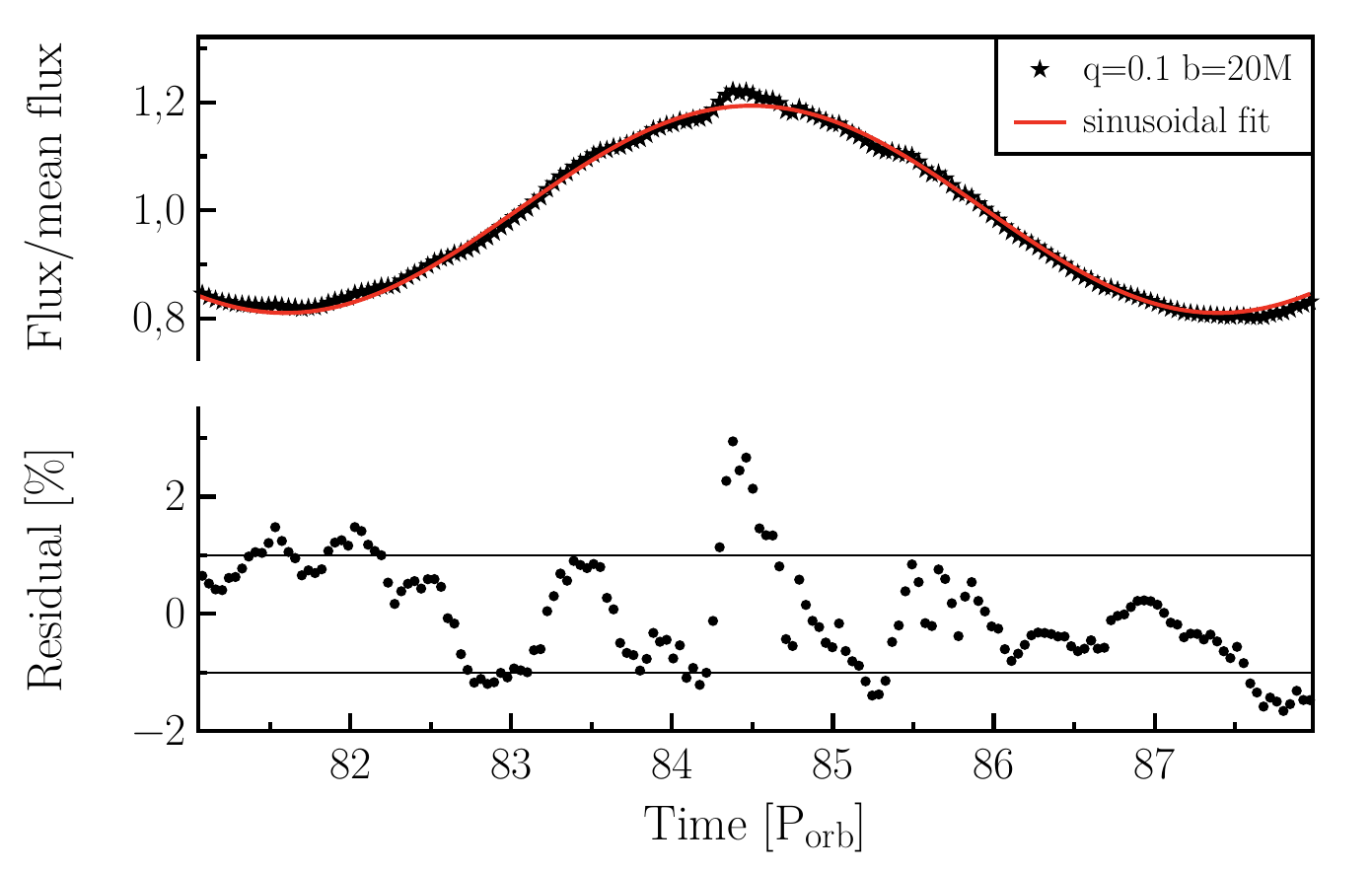}
\includegraphics[width=\columnwidth]{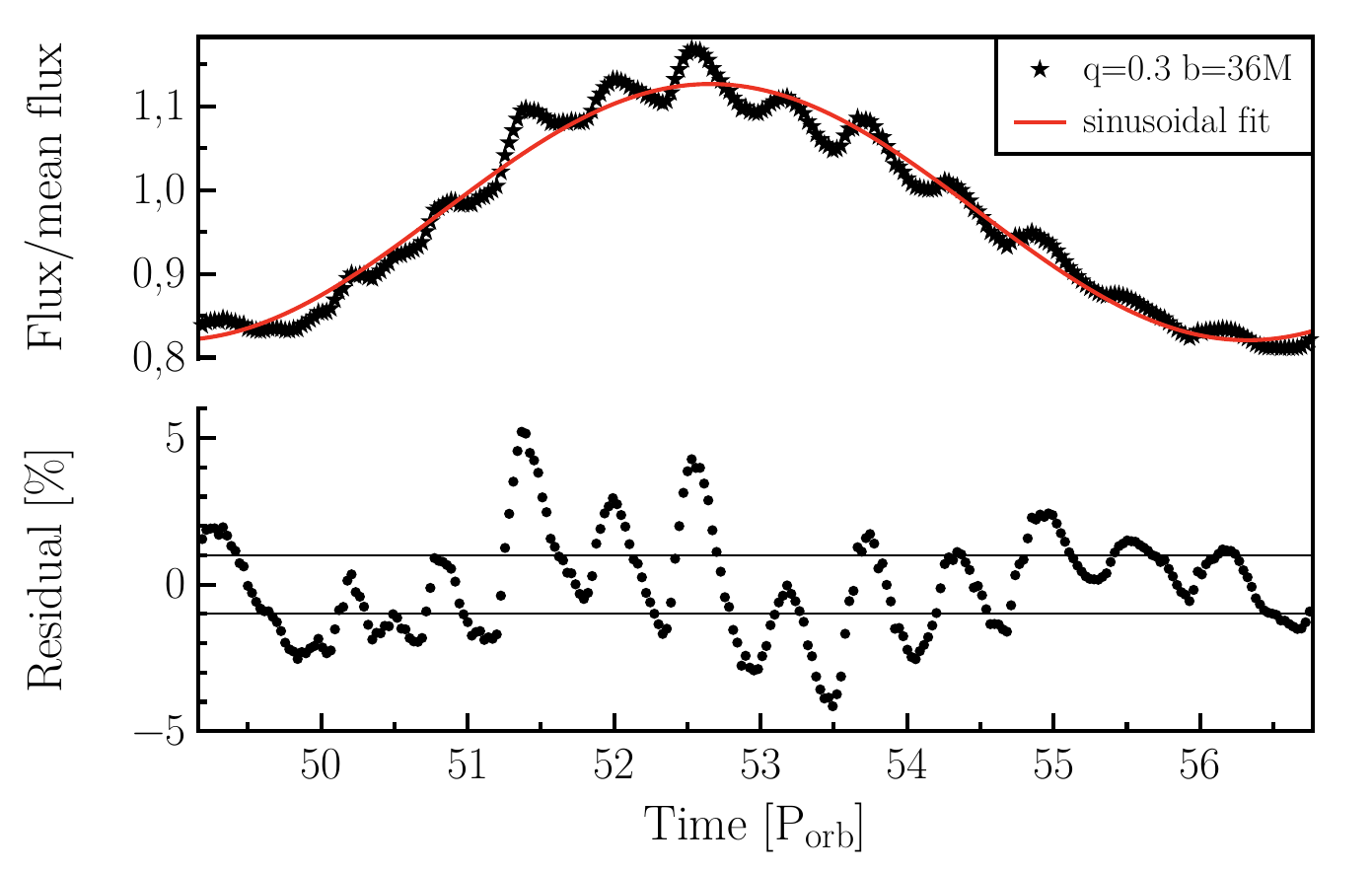}
\caption{{Lightcurves obtained from the fluid simulations fitted with a sinusoidal function (see Eq.~\ref{eq:model}) and fit residuals indicated in $\%$, for $q\,{=}\,1$, $b\, {=} \, 20$~M (top panel), $q\,{=}\,0.1$, $b\, {=} \, 20$~M (middle panel), and $q\,{=}\,0.3$, $b\, {=} \, 36$~M (bottom panel).
The horizontal lines indicate the $1\%$ level.
All LCs have been computed for an inclination of $70^\circ$.
}}
\label{fig:fitsin}
\end{figure}

{For this comparison, shown in Fig.~\ref{fig:fitsin}, we have considered three runs varying the mass ratio or the orbital separation, all exhibiting the lump feature: $q\,{=}\,1$, $b\, {=} \, 20$~M (top panel), $q\,{=}\,0.1$, $b\, {=} \, 20$~M (middle panel) and $q\,{=}\,0.3$, $b\, {=} \, 36$~M (bottom panel).
We fit the LCs originating from the fluid simulation{s} considering the main and longer-timescale modulation, using the simplified toy model, over one lump period to focus on the LC shape.
Fit residuals are shown and expressed in $\%$.
The simplified toy model recovers the main LC characteristic shape, with the residuals remaining mostly below $1\%$ with some points reaching up to $5\%$.
}

{Residuals also show some sinusoidal substructures on shorter timescales than that of the lump for the $q\,{=}\,1$, $b\, {=} \, 20$~M and $q\,{=}\,0.3$, $b\, {=} \, 36$~M cases.
This could justify the use of the toy model with the shorter, additional frequency, if the temporal resolution allows it.
The secondary frequency is either compatible with the spiral arm frequency ($q\,{=}\,1$, $b\, {=} \, 20$~M, top panel) or with the binary-lump beat ($q\,{=}\,0.3$, $b\, {=} \, 36$~M, bottom panel).
We can expect the relative importance of the modes to depend on the disk thermodynamics and on various parameters: larger spiral-arm modulations being hosted in more inclined and closer BBHs while perfectly face-on systems should only exhibit the beat modulation.}\\

{Overall, the simplified toy model reproduces well the main characteristics of these simulated data from the CBD EM variability. 
On top of the lump, the imprint from a second modulation can be visible in the residuals depending on the system's parameters and on the data quality.
In this case the toy model could be used with the second period, $P'$ in Eq.~\ref{eq:model}, corresponding either to the beat or to the spiral-related one.}

\section{Implications in the context of BBH detection}
\label{sec:discu}

Here we are presenting a few ways that this simplified toy model could be used in the context of BBH detection or identification{, discussing timescale constraints and estimating the detectability of the LC modulations in the optical band with the Vera Rubin Observatory (VRO)}.

\subsection{Timescale limitations and the circular-orbit hypothesis}

{
On their way to merger, accreting BBHs possibly go through two phases: (i) slow inspiral, when the timescale of reference (e.g. the lump period) can be considered as constant over the exposure, cadence, and observation times and
(ii) faster inspiral, during which this same period decay (in the pre-decoupling epoch, \citealt{tang_late_2018}, \citealt{dittmann_decoupling_2023}) is no longer negligible over these observation-related timescales.
As we consider circular orbits for the BBH, it categorizes this study, implicitly, in (an asymptotical version of) the slow inspiral phase.
Here, we present the consequences of the inspiral on the modulation periods.
\begin{figure}
\includegraphics[width=\columnwidth]{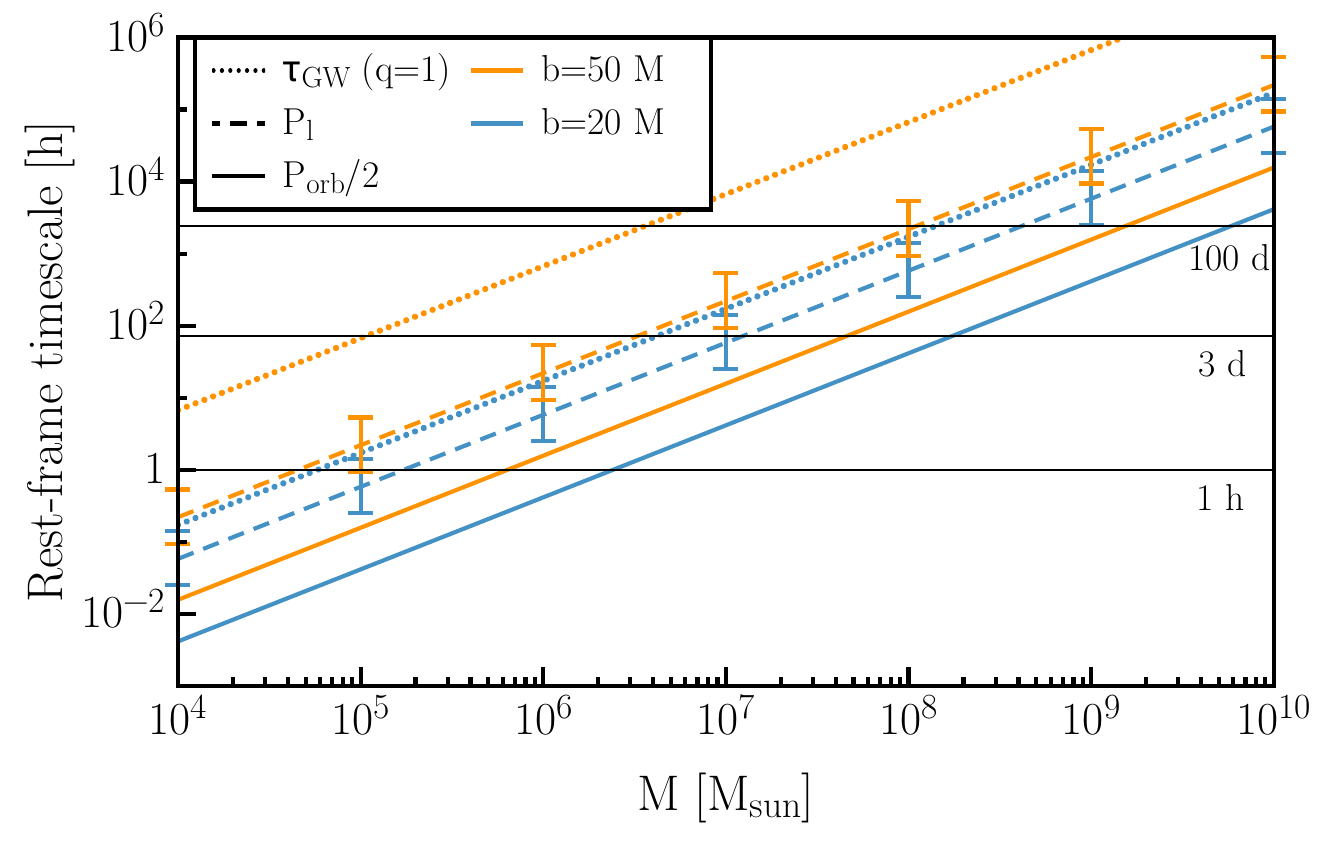}
\caption{Timescales of interest{:} semi-orbital period {(full line)}, lump period {(dashed line) and time to merger, denoted $\tau_\mathrm{GW}$ (dotted line),} for the EM detection of a BBH CBD, as a function of $M$, for $q\, {=} \,1$ and two values of $b$.}
\label{fig:Plumporb}
\end{figure}
For reference, Fig.~\ref{fig:Plumporb} illustrates the evolution of 
the three timescales of interest here, namely $\mathrm{P_{orb}}/2$ {(or the beat period, which is of the same order)}, $\mathrm{P_{l}}$, which are the two modulation timescales to sample, and the time to merger, for $q\, {=} \,1$ as a function of $M$.
As a guide, the time it takes for $P_\mathrm{l}$ to decrease by a factor $X$ (such that the new period is $X\, P_\mathrm{l}$ with $X\, {<}\, 1$), denoted  $\tau_{\mathrm{circ},X}$ is given, 
by the simple formula
\be
\tau_{\mathrm{circ},X} \, {=} \, ( 1 -X^{8/3}) \, \tau_\mathrm{GW},
\ee
with $\tau_\mathrm{GW}$ the initial time to merger \citep{peters_gravitational_1964}; which corresponds, in the pre-decoupling regime, to a decrease of $b$ of $(1-X^{2/3})$.
This timescale is independent of the mass ratio.
For illustration purposes, $\tau_{\mathrm{circ},0.8} \, {\approx} \, 0.45  \tau_\mathrm{GW}$; over this time, $b$ decreases by $14 \%$.
If the integration time approaches $\tau_{\mathrm{circ},X}$ with $X$ such that the period decay is not negligible, the toy model and the analysis method should account for the decaying period.
As an example, Fig.~\ref{fig:Plumporb} shows that for $M\,{\gtrsim}\, 10^8 \mathrm{M_\odot}$, $\tau_{\mathrm{circ},0.8}$ approaches typical optical LCs timescales (hundreds of days to years, \citealt{graham_systematic_2015}) for relativistic separations, in which case the inspiral might need to be taken into account.
}
\\

{Another aspect tangential to the toy model and its applications is the following.
In order to detect a modulation, the instruments we have access to must be able to sample it.
Meanwhile, the characteristic periods scale as $M$ at fixed separation (in gravitational radii, which is the relevant units for the dynamics), so modulations in lower-mass systems occur on shorter timescales.
For example, the LSST, which surveys the entire sky every $3$~days, will not be able to properly sample the lump period for systems with $M\, {\ll}\, 10^7 \mathrm{M_\odot}$ at relativistic separations (Fig.~\ref{fig:Plumporb}).
}

\subsection{Relevance of the circumbinary contribution to the variability of quasars} 
\label{sec:binarity}

One of the interest{s} of this model is that it can be used to test the viability of the BBH hypothesis for quasars exhibiting variability in the {UV/optical} band.
Let us check whether the modulation periods{, shown in Fig.~\ref{fig:Plumporb},} are already accessible with current detectors or even in archival data.
As {mentioned above}, in the optical band, where the energy spectrum of the most massive ($M\, {\ge} \, 10^8\,  \mathrm{M_\odot}$) systems' CBD peaks, ${\sim}100-1000$~days-long variability patterns have been reported in various studies dedicated to similarly heavy BBH candidates (e.g. \citealt{graham_possible_2015}, \citealt{liu_supermassive_2019}).
From Fig.~\ref{fig:Plumporb} these timescales fall into the range of the lump period for $b\, {=} 50$~M ($M\, {=} \, 10^8\,  \mathrm{M_\odot}$) or $b\, {=} 20$~M ($M\, {=} \, 10^9\,  \mathrm{M_\odot}$).
We applied a similar reasoning in \cite{foustoul_catalogue_2025} to derive the potential separation from the observed mass and periodicity.
Similar or better sampling {of the variability periods} can be expected with recent and future EM facilities such as
 the Large Synoptic Survey Telescope\footnote{https://www.lsst.org/about} (LSST) of the {VRO}.
\\

\begin{figure}
\includegraphics[width=\columnwidth]{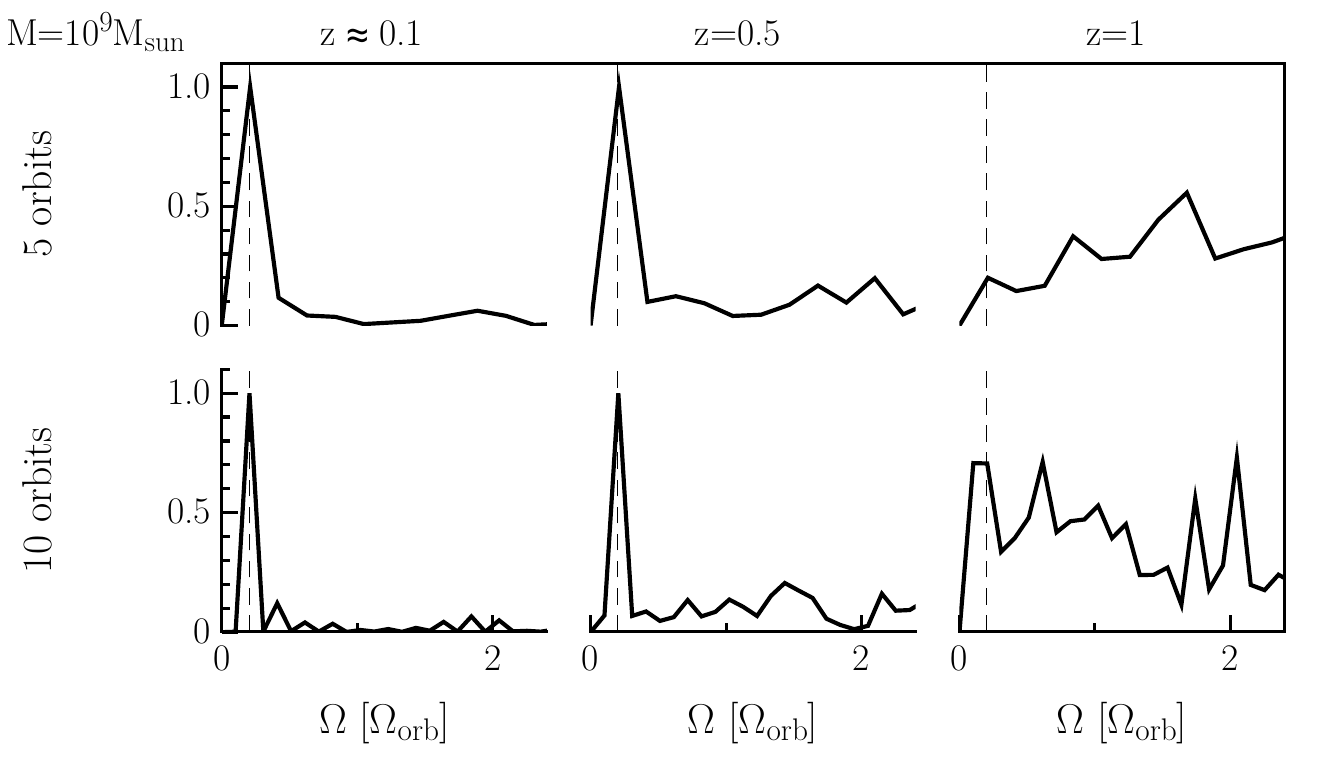}
\caption{
{Result of the Fourier transform onto synthetic LCs, in amplitude normalized by the maximal amplitude, for $b\, {=} \, 20$~M, $M\, {=} \, 10^{9} \, M_\mathrm{\odot}$, for $\gamma_\mathrm{l} \, {=} \, 0.1$, $\gamma_\mathrm{s} \, {=} \, 0.02$, for $z \, {\in} \, \{0.1,0.5,1\}$. The vertical dashed line indicates the lump orbital frequency.}
}
\label{fig:FFT_VRO}
\end{figure}

{As an example, we study the possibility to detect the LC modulations with the VRO, using the same method as in \cite{cocchiararo_electromagnetic_2024}, assuming spiral arm and lump perturbations.
We choose a set of parameters consistent with a CBD peak frequency in the G-band ($[5.43,7.5]\times 10^{14}$~Hz): $b\, {=} \, 20$~M, $M\, {=} \, 10^{9} \, \mathrm{M}_{\odot}$, and add gaussian noise to the LC, with a variance equal to the $5{-}\sigma$ sensitivity flux, i.e. $2.32 \times 10^{-15} \, \mathrm{erg \, cm^{-2} \, s^{-1}}$ (\href{https://pstn-054.lsst.io/PSTN-054.pdf}{https://pstn-054.lsst.io/PSTN-054.pdf}).
Other radiation components are neglected.
We considered a rather pessimistic case for the modulations' amplitude, with $\gamma_\mathrm{l} \, {=} \, 0.1$, $\gamma_\mathrm{s} \, {=} \, 0.02$ (resulting in a $\pm 4\%$ dominant modulation, see Fig.~\ref{fig:diffgammal}).
We consider three values for the distance\footnote{Starting from the LC at $z \, {\sim} \, 0.1$ (Fig.~\ref{fig:diffgammal}), we re-scale the flux as a function of the distance as $F\, {\varpropto }\, D^{-2}$ but, for simplicity, we do not apply the frequency redshift, as it would require a fine-tuning of the mass or the separation to maintain the CBD peak frequency at the same value, in the G-band.}, $D \, {=} \, 500$~Mpc ($z \, {\sim} \, 0.1$), $D \, {=} \, 2.9$~Gpc ($z\, {=} \, 0.5$), and $D \, {=} \, 6.8$~Gpc ($z\, {=} \, 1$) and perform Fourier transforms onto limited timescales of $5$ and $10$ BBH orbits, with the orbital period ${\approx}34$~d. 
For this system's parameters, our circular-orbit model is still reasonable here because, over one year of observation, the period decrease due to GW emission only is minor (around $15\%$, for $q \, {=}\, 1$, less if $q \, {<} \, 1$; $b$ decreases by $10\%$). 
The resulting Fourier amplitude, normalized by the corresponding maximal amplitude, is shown in Fig.~\ref{fig:FFT_VRO}.
For a five-orbits integration, the lump modulation is visible up to $z \, {\sim} \,0.5$\footnote{{
In comparison with the $q\, {=}\, 1$, $e\, {=}\, 0$ case of \cite{cocchiararo_electromagnetic_2024}, we obtain a better detectability of the lump.
Indeed, although the modulation's normalized amplitude is similar, by choosing a BBH mass of $10^9 \mathrm{M_\odot}$ at relativistic separation instead of $10^6 \mathrm{M_\odot}$ at large (${\sim} 10^4 \mathrm{r_g}$) separation, our mean optical luminosity is ${\sim} 4\times 10^{-12} \mathrm{erg  \, cm^{-2} \, s^{-1}}$ while theirs is ${\sim} 10^{-13} \mathrm{erg  \, cm^{-2} \, s^{-1}}$ at similar redshift (see their Fig.~6), enhancing the signal-to-noise ratio.
}
}
; for a ten-orbits integration (${\sim} 1$~yr, one tenth of the planned lifetime of VRO), its detection up to $z \, {=}\, 0.5$ is increasingly convincing with a ratio of $10$ with respect to nearby frequencies -- that is $100$ if expressed in power instead (e.g. \citealt{noble_circumbinary_2012}) -- while for $z \, {=}\, 1$ the noise dominates (with the maximal being located outside of the displayed frequency range).
Our results suggest that only a larger temperature perturbation could allow for a detection at $z \, {>}\, 1$.
Meanwhile, the spiral-related modulation is not detectable; a similar result would be obtained for the binary-lump beat with the same modulation's amplitude.
We specify that we used the synthetic LC here, but the toy model could be used instead to perform a similar exercise but this time to determine the minimal modulation's amplitude allowing for a detection at a given redshift.}

\subsection{Estimating the LC modulation period for a GW event}

A more direct use of this toy model is to help {in} the search {for EM} counterpart of GW sources detected by LISA. 
While LISA will detect the GW signal {from massive BBHs} prior to the merger, the time {available} to find the source will range from months for the best cases to days and even hours before merger \citep{mangiagli_observing_2020}.
This means that we need a fast method {capable of providing} the optimal energy band for detection as well as the period (Eq.~\ref{eq:periods}) we are looking for.
Our toy model will be able to {provide these once} the GW detection has given constraints on $M$, $q$, and time to merger, {from which $b$ is derived}, hence
allowing us to focus our observations with the observatory having the best chance to detect the variability.

Even more importantly, our toy model could also be used to {investigate} the past history of the modulation and thus {allowing us} to extend our search for the EM counterpart 
to real-time search but also using archival data of the region of interest (see \citealt{xin_identifying_2024} on using LSST archival data to search for a LISA counterpart).
Such search could be automated as  the binary separation can be extrapolated back in time{,} and our toy model can estimate what was the periodicity for any archival observation 
data of the region.

 {This} approach can be taken a step further by looking at what {the current} characteristics of a {potential future} LISA {source} could {be}. This would allow us to have,
 by the launch of LISA, a catalog of potential BBHs from which we could pick the best target once LISA starts detecting GWs. 
 Having such {a} catalog would speed up the potential detection of the EM counterparts of the LISA detection. 

\section{Conclusions}
\label{sec:ccl}

In this paper, we have constructed an analytical non-axisymmetric CBD temperature distribution {of BBH systems that incorporates commonly reported features of the CBD such as} spiral arms and the {\lq}lump{\rq}. 
Once ray-traced back to the observer {in a BBH spacetime}, these non-axisymmetries {produce sinus-like modulations} 
 in the UV/optical band for BBH masses $M \, {=} \, 10^{4} \, \mathrm{M_\odot}$ to $10^{10}\, \mathrm{M_\odot}$.
We proposed a simple toy model that reproduces the LC variability {from ray-traced LCs from the aforementioned temperature distribution. It also reproduces the main characteristics of LCs from 2D GRHD simulations under the assumption of adiabatic heating/cooling, apart from the binary-lump beat, which arises from the simulation and is intrinsically dynamical. 
However, the latter also produces a sinusoidal signal and could be integrated to the toy model}.

The advantage of this model lies in its simplicity.
Indeed, it could be used to fit optical/UV quasar LCs, retrieving the semi-orbital and/or lump period.
Alternatively, in the LISA era, it could rapidly provide an estimate of the optimal energy band and variability timescale associated with the CBD thermal emission.
This variability falls into the band of the {VRO}, the Hubble Space Telescope\footnote{https://science.nasa.gov/mission/hubble/}, the Zwicky Transient Facility (ZTF)\footnote{https://www.ztf.caltech.edu}, and the future UV Explorer\footnote{https://www.ipac.caltech.edu/project/uvex} and Habitable Worlds Observatory\footnote{https://science.nasa.gov/astrophysics/programs/habitable-worlds-observatory/}.
{We showed that, for $b\, {=} \, 20$~M, $M\, {=} \, 10^{9} \, \mathrm{M}_{\odot}$, and in the absence of other contaminating radiation components, even a mild temperature perturbation from the lump would be detectable with VRO up to $z \, {=} \, 0.5$ with six months to one year of observation.
We should keep in mind, though, that detecting a variability is not a definitive test for BBHs because single BHs could also produce similar variabilities \citep{varniere_can_2025}.
}

{We have explicitly focused on the CBD thermal emission to take advantage of the progress made in the past years on the CBD dynamics and emission properties (see \citealt{gold_relativistic_2019}, \citealt{dorazio_observational_2023} and references therein).
Other sources of emission, including the thermal emission from the individual disks to begin with, would add up to the CBD contribution, as well as their variability, albeit at higher frequency (e.g. \citealt{gutierrez_electromagnetic_2022}).
In the case of quasars, obscuration by a dusty torus could drastically reduce the observed flux, especially at high inclination where the relativistic boost is actually maximal.
These components will be included in future works.}

While this work was developed, a tool to compute the EM variability of eccentric BBHs has been proposed by \cite{dorazio_fast_2024}.
They focus on accretion-rate related variability affecting the disk emission (on which relativistic Doppler effect is applied), while we explicitly focus on relativistic Doppler effect acting directly on non-axisymmetric temperature distribution in the CBD to propose a toy model. Thus, we note the complementarity of both approaches.

\section*{Acknowledgements}

RMR thanks A. Coleiro, P-A Duverne, F. Cangemi, V. Foustoul, A. Dittmann, D. D'Orazio and S. Lescaudron for useful discussions.
RMR acknowledges funding from Centre National d'Etudes Spatiales (CNES) through a postdoctoral fellowship (2021-2023). 
RMR has received funding from the European Research Concil (ERC) under the European Union Horizon 2020 research and innovation programme (grant agreement number No. 101002352, PI: M. Linares).
This work was supported by CNES, the LabEx UnivEarthS, ANR-10- LABX-0023 and ANR-18-IDEX-000, and by the \lq Action Incitative: Ondes gravitationnelles et objets compacts{\rq} and the Conseil Scientifique de l'Observatoire de Paris. 
This work was also supported by the Programme National des Hautes Energies (PNHE) of CNRS/INSU, co-funded by CNRS/IN2P3, CNRS/INP, CEA and CNES.
The numerical simulations we have presented in this paper were produced on the platform DANTE (AstroParticule \& Cosmologie, Paris, France) 
and on the high-performance computing resources from GENCI - CINES (grant A0100412463) and IDRIS (grants A0130412463 and A0150412463).

\section*{Data Availability}

The data that support the findings of this study are available from the corresponding author, R.M.R, upon request, and will also be part of a data release in 2025\footnote{Which will be 
available for download at
\url{https://apc.u-paris.fr/~pvarni/eNOVAs/LCspec.html}}.

\appendix

\section{Residual eccentricity at small separation}
\label{app:ecc}

\begin{figure}
\includegraphics[width=\columnwidth]{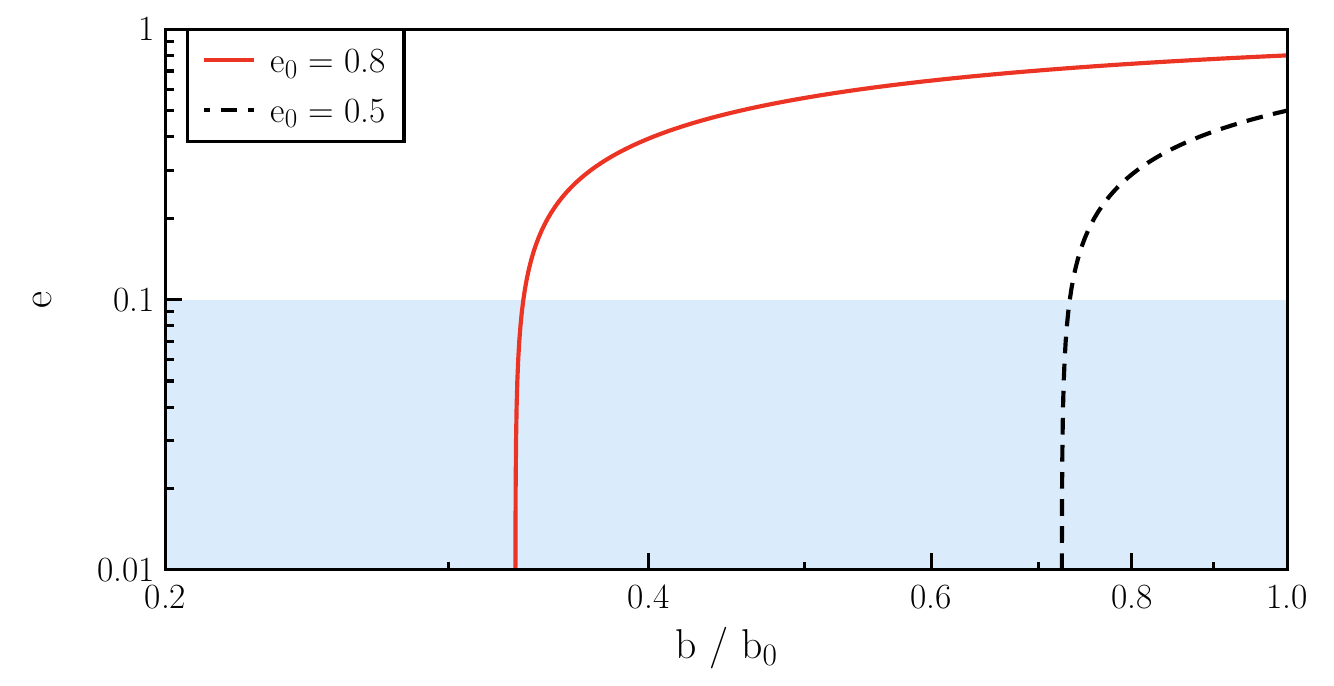}
\caption{{Eccentricity as a function of the orbital separation, for two initial eccentricities: $e_0\, {=}\, 0.8$ (red, full line) and $e_0\, {=}\, 0.5$ (black, dashed line).}}
\label{fig:ecc}
\end{figure}

{Lump disappearance has been found for binaries with eccentricity $e \, {>} \, 0.1$ in 2D locally isothermal CBDs by e.g. \citealt{siwek_preferential_2022} while it has been reported for $e \, {=} \, 0.9$ (and $q \, {=} \, 1$) in 3D self-gravitating CBDs by \cite{cocchiararo_electromagnetic_2024}.
Thus, while the overall impact of the binary's eccentricity onto the lump formation is not settled yet, these studies agree when the orbit is circular.
Hence, we check whether supermassive BBHs at relativistic separations, on which we focus here, have a residual eccentricity compatible with the lump's presence.
For this, we first compute the semi-major axis as a function of the eccentricity, using the post-Newtonian Eq.~5.48 of \cite{peters_gravitational_1964}.
We consider an initial eccentricity compatible with the equilibrium value from binary-disk interaction, i.e. $e_0 \, {=} \, 0.5$ (\citealt{dorazio_orbital_2021}, \citealt{zrake_equilibrium_2021}, \citealt{siwek_orbital_2023}; see also the observational study of \citealt{murray_accreting_2025} on binary stars), or even higher, i.e. $e_0 \, {=} \, 0.8$.
For this first application, we neglect any mechanism increasing the eccentricity.
Figure~\ref{fig:ecc} shows that the eccentricity reduces from $e_0\, {=} \, 0.5$ (resp. $e_0\, {=} \, 0.8$) to $e\, {<} \, 0.1$ as the semi-major axis decreases from an arbitrary initial value $b_0$ to $0.73 \, b_0$ (resp. $0.33\, b_0$) only.
This illustrates the efficiency of GW-driven eccentricity damping and that BBHs reaching relativistic separations would likely have $e \, {<} \, 0.1$, thus being compatible with the lump's presence.}

{If we assume an eccentricity equilibrium obtained through binary-disk interaction, this same mechanism could be at work and counteract the GW-driven eccentricity damping.
Thus, let us compare the timescales of eccentricity increase and GW-driven damping.
For the former, \cite{siwek_orbital_2023} found a few Myr for a binary accreting at its Eddington rate with a radiative efficiency $\epsilon\, {=}\, 0.1$.
For the latter, we compute it as $\tau_\mathrm{GW,e} \, {\equiv} \, e / (\mathrm{d}e / \mathrm{d}t)$ using Eq.~5.42 of \cite{peters_gravitational_1964} and find}
\be
\tau_\mathrm{GW,e} \, {\approx} \, 0.8 \left( \frac{ b}{100 \, \mathrm{r_g}} \right)^4  \frac{(1+q)^2}{q} \left( \frac{M}{10^6 \mathrm{M_\odot} } \right)  \mathrm{yr},
\ee
{which is orders-of-magnitude shorter than the eccentricity increase timescale found by \cite{siwek_orbital_2023}, at relativistic separations, even for more massive ($10^9 \mathrm{M_\odot}$) BBHs like PTA sources.
This result agrees with a low residual eccentricity found by \cite{zrake_equilibrium_2021} when supermassive BBHs enter the LISA band.
We conclude that supermassive BBHs at relativistic separations likely have a low eccentricity $e \, {<} \, 0.1$, compatible with the lump.}



\bibliographystyle{mnras}
\bibliography{Zotero.bib} 






\label{lastpage}
\end{document}